\documentclass[lettersize,journal]{IEEEtran}
\usepackage{amsmath,amsfonts}
\usepackage{array}
\usepackage[caption=false,font=small,labelfont=sf,textfont=sf]{subfig}
\usepackage[font=small,labelfont=bf]{caption}
\usepackage{textcomp}
\usepackage{stfloats}
\usepackage{url}
\usepackage{verbatim}
\usepackage{graphicx}
\usepackage{cite}
\usepackage{balance}
\usepackage{algorithm}
\usepackage{algpseudocode}
\usepackage{enumitem}
\usepackage[english]{babel}
\usepackage{hyperref}
\addto\extrasenglish{%
}
\makeatletter
\newcommand*{\rom}[1]{\expandafter\@slowromancap\romannumeral #1@}
\makeatother
\graphicspath{{figs/}{figures/}{pictures/}{images/}{./}} 

\begin{document}
\title{NOVA: A visual interface for assessing polarizing media coverage}
\author{
Keshav Dasu~\IEEEmembership{Member, IEEE}, Sam Yu-Te Lee, Ying-Cheng Chen, Kwan-Liu Ma, \IEEEmembership{Fellow, IEEE}
\thanks{Keshav Dasu, Sam Yu-Te Lee, Ying-Cheng Chen, and Kwan-Liu Ma are with the Department of Computer
Science, University of California at Davis, One Shields Ave., Davis, California
95616. E-mail: {kdasu, ytlee, ycchen, klma}@ucdavis.edu}
}

\markboth{IEEE TRANSACTIONS ON VISUALIZATION AND COMPUTER GRAPHICS}
{Keshav Dasu, Sam Yu-Te Lee, Ying-Cheng Chen, Kwan-Liu Ma, \MakeLowerCase{\textit(et al.)}:NOVA: A visual interface for assessing polarizing media coverage }%

\maketitle
\begin{abstract}
 Within the United States, the majority of the populace receives their news online. U.S mainstream media outlets both generate and influence the news consumed by U.S citizens. Many of these citizens have their personal beliefs about these outlets and question the fairness of their reporting. We offer an interactive visualization system for the public to assess their perception of the mainstream media’s coverage of a topic against the data. Our system combines belief elicitation techniques and narrative structure designs, emphasizing transparency and user-friendliness to facilitate users' self-assessment on personal beliefs. We gathered $\sim${25k} articles from the span of 2020-2022 from six mainstream media outlets as a testbed. To evaluate our system, we present usage scenarios alongside a user study with a qualitative analysis of user exploration strategies for personal belief assessment. We report our observations from this study and discuss future work and challenges of developing tools for the public to assess media outlet coverage and belief updating on provocative topics.
\end{abstract}


\vspace*{-0.2cm}
\section{Introduction}
\noindent In the current age, written news media is a combination of content that is either automatically generated by news algorithms~\cite{diakopoulos2019automating} or long thematic pieces curated by journalists to provide deeper analysis and thorough explanations to the public. 
Simultaneously, many methods and different stages of the journalistic process inject biases into a news article. 
How an article is written and framed~\cite{hamborg2019automated} can strongly impact one's opinions and perspectives on issues.
Media bias and its effects are widely documented~\cite{wolton2019biased,shu2020fakenewsnet,chen2020analyzing,spinde2020enabling} and known to the general public.

Articles published by outlets they perceive as ``biased'' are often dismissed as ``fake news''.
The continued dismal of other perspectives can gradually form an echo chamber as people unintentionally insulate themselves within their personal beliefs~\cite{gillani2018me}.

Recent developments in Natural Language Processing have sparked an increased effort to rethink how we analyze media bias. 
Social science and psychology researchers have been calling for an interdisciplinary approach with computer science to assess news media content~\cite{hamborg2019automated}. 
Previous works have applied machine learning models on annotated datasets to identify certain types of media bias~\cite{spinde2021neural, hube2019neural, hube2018detecting, chen2020analyzing}, but they are susceptible to performance issues~\cite{dallmann2015media, spinde2021you}. 
A key reason for that is the difficulty of building a high-quality bias dataset. Readers are unable to agree on what’s biased due to factors like background knowledge, political ideology, or even the Hostile Media Effect in psychology~\cite{spinde2020enabling}. 
Even with recent advances in Large Language Models, it is still challenging to resolve the issue among readers.

To address these issues, we propose to create an environment where people externalize their beliefs on media biases and find evidence to support or disprove them.
This way, belief conflicts among people are effectively avoided.
To guide user exploration within such an environment, we incorporate the interactive presentation narrative structure~\cite{segel2010narrative} by dividing the environment into three stages, each with a clear task and set of instructions.
Previous works have reported that ``the interpretations of the text can be multiple and they depend on the personal background knowledge, culture, social class, religion, etc. as far as what is normal (expected) and what is not are concerned''~\cite{balahur2013sentiment}. Recognizing the complexity of beliefs, we choose to guide user exploration with less contentious ``signals'', i.e., sentiment and named entities, building upon prior works that have shown the viability of combining them to identify framing characteristics~\cite{balahur2013sentiment, abbar2013real, Park2011, vanRel2020}. 
Note that in our work, the entities and sentiments direct the users to potential media biases, while the users have full control over which direction to go, and whether an actual bias is found.

We are interested in exploring the relationship between one’s personal beliefs about news media outlets and how those outlets behave. We take into account the key findings in other fields and introduce an alternative approach to explore and assess if one’s perception of a media outlet aligns with how that outlet reported on a topic. To achieve this, we introduce a system for News Outlet Visual Assessment (NOVA).
\textit{NOVA is a visual interface designed to facilitate the self-assessment of personal beliefs on news media coverage biases.} 

Our target audience is a subset of the general public. We cater to individuals who either encounter or actively seek out news media articles and content (e.g., NYT, CNN, etc) fairly regularly (i.e., at least monthly). We walk through an example workflow to demonstrate how NOVA can facilitate assessing media outlet coverage as it relates to one’s personal beliefs. Additionally, we present a user study with a qualitative analysis of user exploration strategies and challenges for personal belief assessment. 
Our contributions are as follows:
\begin{itemize}
  \setlength\itemsep{0em}
    \item We introduce NOVA, a multi-stage system that combines presentation narrative structure and belief elicitation for the general public to self-assess their personal belief on news media coverage.
    \item We identify lessons learned from designing and evaluating an interactive visualization system for belief elicitation and assessment and prospective research directions. 
\end{itemize}

\section{Related Works}
\noindent Our work lies at an intersection of text analysis, as it pertains to news-related content, and the assessment of an individual’s belief or preferences and illustrating those effects. 
In this section, we review visual analytic systems with a focus on news media, personal belief visualization systems,  and visualization works that foster user engagement with complex data.

\vspace*{-0.3cm}
\subsection{News Visual Analytic Systems}
We review visual analytic systems with a focus on evaluating news media.
Particularly, systems that either support domain experts in their meta-analysis or those that offer the public a means to see more than what is written. 

The COVID-19 pandemic itself produced many research works~\cite{Zhang2021covid,Kong2018} assessing how information was disseminated and various phenomena that happened in that period of history. 
Zhang et al.~\cite{Zhang2021covid} examined the visualizations produced during that time and assessed how visualizations were used in crises to disseminate information. 
Kong et al.~\cite{Kong2018} studied how visualization titles can introduce subtle slants, which could bias a viewer’s perception of the content.
A goal of NOVA is to enable the public to assess the written media content from mainstream U.S. media outlets, rather than the visualization content these outlets may also have put out. 
Additionally, we are interested in studying the relationship between an individual’s perception of news outlets and the content the outlets produce.
One method of understanding these biases is to visualize this relationship using sentiment analysis.

There are several visual analytic systems~\cite{Ilyas2020,hamborg2021newsalyze,urologin2018sentiment} that primarily use sentiment to examine articles. 
Ilyas et al.~\cite{Ilyas2020} use the mean sentiment score of daily tweets to summarize a daily sentiment. 
Based on the sentiment score fluctuation some extremely negative or positive topics are identifiable.
Hamborg et al.~\cite{hamborg2021newsalyze} use a weighted sum rather than a mean to summarize the sentiment score of an individual.
However, a common problem these works face is when statistics are used to summarize the overall sentiment of a collection of articles, polarizing and neutral ones are usually indistinguishable. 
For example, it is fairly common to group a collection of related articles as either one topic or event and assign an overall sentiment to this single collection. 
For a polarizing collection, the impact of positive articles and negative articles that are contained typically negate each other. 
Other works~\cite{saleiro2015popmine, diakopoulos2010diamonds} try to use the ratio of positive and negative scores, yet also succumb to the same issue.
With NOVA, we address this problem by introducing a two-dimensional sentiment score for any collections. 
Recognizing that an overly complex sentiment score could confuse the general audiences, our design focuses on presenting this content in an interpretable manner that is easy for people to comprehend. 

\begin{figure*}[t!]
  \centering
  \includegraphics[keepaspectratio, width=0.95\textwidth]{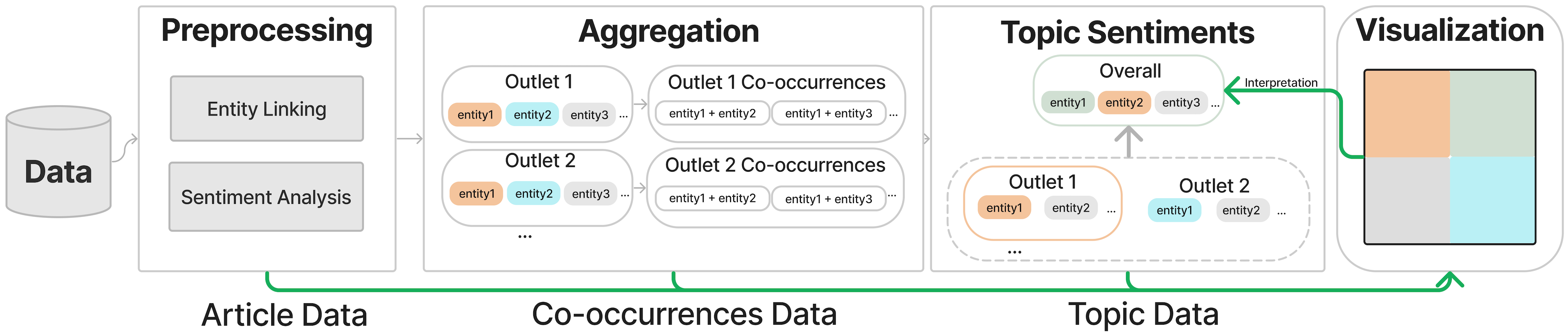}
  \caption{The data transformation process of NOVA. Collected news articles were preprocessed with entity linking and sentiment analysis. Then articles are aggregated by entities and further aggregated by co-occurrences to represent topics. Sentiment scores are generated with descriptive statistics for each topic. The preprocessed article data, co-occurrences data, and topic sentiment data are all stored on a server and requested from the front end. Through user interaction, the aggregated sentiment type of each entity is categorized as neutral, positive, negative, or mixed. Green lines indicate data communication between the server and the front end. }
\vspace*{-0.35cm}
  \label{fig: transformation}
\end{figure*}

\vspace*{-0.3cm}
\subsection{Personal Belief Visualization}\label{sec:personal_bias}
Researchers in information visualization have recognized that people's prior beliefs about the data can influence how they interpret visualizations~\cite{xiong2022seeing}. 
Belief-driven visualizations are thus designed to elicit and visualize people's beliefs alongside data-driven visualizations for improved data analysis or decision-making~\cite{mahajan2022vibe, kim2017explaining}.
Previous works commonly define ``belief'' as a mental presentation that people can express using numerical or categorical parameters, e.g. Bayesian probabilistics~\cite{kim2020bayesian, kim2019bayesian}.
However, people's perception of news media coverage is difficult, if not impossible, to quantify.
In NOVA, we rely on the construction of visualization via arranging visual objects to elicit people's beliefs about news media coverage, instead of asking people to provide numerical values that reflect their beliefs.

Another important goal in belief-driven visualizations is to encourage belief-updating or overcoming existing biases~\cite{mahajan2022vibe, kim2020designing}.
In particular, Heyer et al.~\cite{heyer2020pushing} did a large-scale experiment to investigate the effects of belief-driven visualizations on people's belief in provocative topics.
They found that people commonly express surprise when their beliefs are contrasted with data-driven visualizations and are willing to reconcile their lack of prior knowledge.
However, they observed insignificant belief updates in their experiment, which they attributed to the high variation of willingness to change beliefs among individuals.
Considering the controversial nature of news media coverage, we believe that it is impractical to expect people to immediately change their beliefs even after seeing direct disproofs from data.
In NOVA, the data-driven visualization is primarily a reference for people to compare their beliefs against, not to represent the truth.

\vspace*{-0.3cm}
\subsection{Visualization For the Public}\label{sec: viz_for_public}
Our system design draws significant inspiration from prior works that foster user engagement with complex data, especially those targeting the general public.
The subspace of narrative visualization offers many insights and techniques to engage the general public. 
Segel et al.~\cite{segel2010narrative} reviewed data stories created by journalists and proposed a framework for narrative visualization, categorizing their genre, tactics, and structures. 
Dasu et al.~\cite{dasu2020sea} similarly applied narrative visualization techniques in a museum exhibition to present complex scientific findings to the public. 
From these works, we identify the utilization of a narrative structure as a key factor in keeping users engaged in a prolonged self-assessment process and design our system accordingly.

Studies from~\cite{ma2019decoding, peck2019data, stokes2022striking} provide guidelines for improving visualization novices' sense-making process.
Ma et al.~\cite{ma2019decoding} investigated the general public's process for decoding a visualization of complex scientific data. 
They identified several design practices that could hinder the user's decoding process.
Peck et al.~\cite{peck2019data} explored factors that may drive attention and trust in rural populations. They found that novice users prefer visualization that enables them to quickly get the gist, instead of charts with rich information for them to discover.
Additionally, as we are dealing with text data, we are interested in the optimal balance between text and graphics.
In a large-scale experiment conducted by Stokes et al.~\cite{stokes2022striking}, they found many counter-norm design practices receive positive feedback. 
For example, they found that heavily annotated charts were not penalized, suggesting a high preference for text among the general public.
A common finding amongst these works is that novice users tend to struggle to decode color~\cite{ma2019decoding, peck2019data, lee2015people}. Many experiments and field studies have shown that color is the most common encoding that confuses users. 
Thus, our use of color in NOVA should be intuitive, simple, and consistent.

\vspace*{-0.4cm}
\section{Data Processing}
\noindent Our project has been gathering news articles from January 2020 -- January 2023 from six mainstream media outlets, namely ABC News, Breitbart, CNN, Fox News New York Times, and Washington Post. 
The data is stored in a Postgres database and hosted on an AWS EC2 server. For each article, the source, title, content, time published, and URL are recorded.
The data transformation process is depicted in~\autoref{fig: transformation}. 
For each article, we first conduct entity linking and sentiment analysis.
Then the result goes through an aggregation stage, where the articles are aggregated by mentioning entities, outlets, or sentiments.
The pre-processing and aggregation are pre-computed and the results are hosted on our server to be queried. 
Our front end takes input from user interaction and conducts a final transformation to visualize the data. 
Below, we describe how we perform entity linking, sentiment analysis, and data aggregation.

\subsubsection{Entity Linking}\label{sec: el}
A \textit{Named Entity} is a word or phrase mentioned in the unstructured text that references real-world objects, such as a person, an organization, or a location. 
\textit{Named Entity Recognition} extracts mentions in text and assigns a \textit{type} to each mention. 
\textit{Entity Linking} further assigns a unique identifier (e.g., Wikipedia page ID or any URI) for each extracted mention, therefore ``linking'' different mentions of the same entity. 
For example, ``Donald Trump'', ``Trump'', and ``President Trump'' are different mentions referring to the same person, which can be recognized by an entity linking model. 

NOVA uses the linked entities to aggregate news articles. 
Each aggregated result represents a topic.
Prior works generate \textit{topics} by incorporating topic modeling techniques, while we use entities to represent topics.
We use named entities because current topic modeling techniques are not suitable for our objectives. 
Topic modeling techniques such as LDA~\cite{chen2015proposal} are known to be uninterpretable for non-experts and therefore misleading~\cite{lee2017humantouch}.
On the other hand, named entities have the advantage of being self-explanatory and the ability to provide context for a topic.
We used the Radboud Entity Linker (REL) model~\cite{vanHulst:2020:REL} for our entity linking task. 
REL detects and assigns a Wikipedia ID to each entity, allowing us to link entities mentioned in different articles. 
Additionally, we intentionally ran REL on a sentence level
to incorporate the subsequent sentence-level target-dependent sentiment analysis.

\subsubsection{Sentiment Analysis}\label{sec: sst}
The sentiment analysis result is core to the identification of media bias in NOVA; therefore, the extracted sentiment needs to be both accurate and interpretative. 
We use NewsSentiment~\cite{Hamborg2021b}, a target-dependent sentiment classification model trained for news articles. 
NewsSentiment runs on top of the named entity linking result described in~\autoref{sec: el} and assigns a sentiment type (positive, negative, or neutral) to each entity in the sentence.
Note that if more than one entity appears in a sentence, each entity sentiment is assigned independently to guarantee accuracy. 

To ensure simplicity and therefore interpretability, we further aggregate the sentence-level results into document-level. 
For each news article, we run NewsSentiment on each sentence in the article that mentions at least one entity and classifies the sentence as being positive, negative, or neutral towards the mentioned entity. 
Then for each entity mentioned in an article, we take $\max(\#pos\_sentences, \#neg\_sentences, \#neu\_sentences)$ as the target-dependent document-level sentiment.

\vspace*{-0.15cm}
\section{NOVA: The First Design}\label{sec: first_design}
\noindent The main objective of NOVA is to (1) allow general audiences to freely assess mainstream media coverage on a variety of topics,
and (2) to provide a platform for people to assess their personal beliefs about mainstream media outlets.
We conducted two rounds of design and evaluation to ensure our systems meet both these objectives.
In the initial design, we derived our design choices from existing works and conducted a pilot study to evaluate the feasibility of the design.
In the second round, we refined the design based on the pilot study results, derived new design considerations, and redesigned NOVA significantly.
Below, we describe our initial design and the pilot study results, then highlight the changes we made in the second round of design.

\begin{figure}[t]
    \centering
    \includegraphics[keepaspectratio, width=0.8\columnwidth]{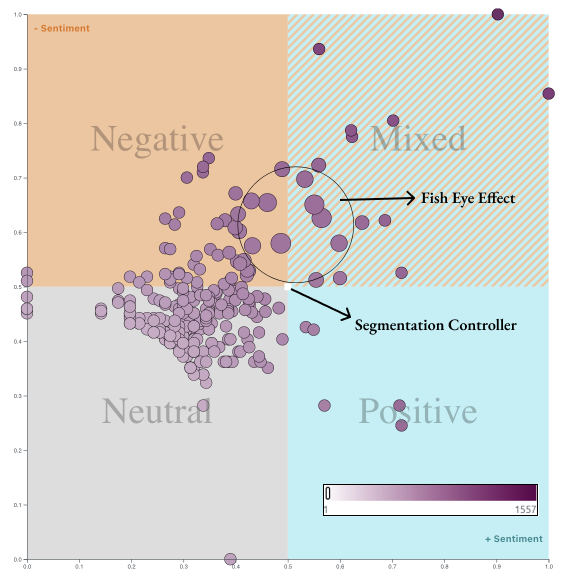}
    \caption{
        Sentiment Scatter Plot. Each dot encodes a topic.
        The color of the dot encodes the number of articles associated with the topic, using a logarithmic color scale.
        Coordinates encode the two-dimensional sentiment of the topic's coverage, using $score_{pos}$ as the x-axis and $score_{neg}$ as the y-axis.
        A segmentation controller divides the scatter plot into four regions: \textit{neutral, positive, negative, and mixed}.
        Topics that fall into each region are classified accordingly, which are used in other parts of the system.
        Finally, a fish-eye effect is added to mitigate the cluttering issue.
    }
    \vspace*{-0.3cm}
\end{figure}\label{fig: sst-scatterplot}

\vspace*{-0.35cm}
\subsection{Visualization Design}
A challenge in visually assessing the coverage of media outlets is to present polarizing topics separately from neutral ones. 
Through our data analysis, we found a way to distinguish these from one another; however, we needed an intuitive and effective way to visualize these results. 
With \textit{Sentiment Scatter Plot}, we address this issue by visualizing a 2D sentiment of a topic where each quadrant implication is easy for our users to understand. 
The Sentiment Scatter Plot serves as a high-level overview of the data and supports users to select topics of interest.
Once a topic is selected, users can then explore the context of the topic by interacting with the \textit{Topic Hive}.
Both the Sentiment Scatter Plot and Topic Hive are designed to be readily decipherable and intuitive to use, taking into account the visual literacy of general audiences.
We first introduce the visual encoding of both visualizations and then describe how they are used in different stages of the system.

\subsubsection{Sentiment Scatter Plot}\label{sec: sst_scatterplot}
A problem with using descriptive statistics (e.g., mean) to measure the sentiment of a collection of articles is that the results for the following two cases are often indistinguishable: (1) when most articles are neutral and (2) when articles have equally high positive and negative scores as the positive and negative scores negate each other numerically. 
However, such topics in actuality are under mixed coverage or polarizing. making them potentially interesting to study further. 
Therefore, we must distinguish these two cases from one another. 

To make this distinction, we treat positive scores and negative scores as two independent variables associated with the coverage of the topic and plot each topic as a dot on a scatter plot (\autoref{fig: sst-scatterplot}). 
We use a scatter plot as the general public is quite familiar with this representation and understands how to interpret it. 
The positive score and negative score of a topic's coverage encode statistics of the associated positive and negative articles, respectively. We use min-max normalization on the number of articles and normalize the scores into [0, 1]. 
Then we plot each topic using $score_{pos}$ as the x-axis and $score_{neg}$ as the y-axis. We further encode the number of articles with a logarithmic color scale to assist users in understanding the underlying statistics. 

To help users understand the meaning of a polarizing topic, we use a \textit{segmentation point} to explicitly divide the scatter plot into four regions: \textit{neutral, positive, negative, and mixed}. 
Topics that fall into each region are classified accordingly. 
The segmentation point controller can be moved by users to adjust the segmentation threshold. 
Previous works commonly used hard-coded thresholds (e.g., 0.33) for dividing positive, negative, and neutral scores. 
The hard-coded threshold may lower trust from users if they find a topic being classified in a way they don't agree with, which is not desirable 
since we aim to create a platform for users to externalize their beliefs.
Finally, we incorporate fish-eye interaction to reduce cluttering, which is a user-controlled focus point for indicating which part of the scatter plot is to be zoomed. A circle around the cursor indicates the area to be zoomed. When the cursor moves towards a particular node, the node will be attracted along the moving direction for an easier selection. 

\begin{figure}[t]
    \centering
    \includegraphics[keepaspectratio, width=0.8\columnwidth]{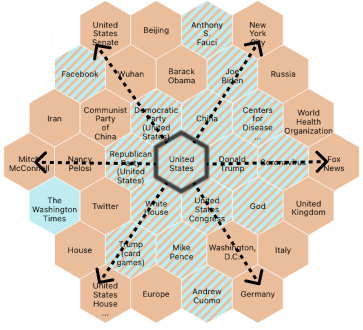}
    \caption{
        Topic Hive first design. Each cell represents a topic. 
        The hive is built around the center cell (United States). 
        Surrounding cells represent the most frequently co-occurring topics with the center cell (topic). 
        Distance of surrounding cells to the center cell encodes the frequency of co-occurrence: closer cells (topics) co-occur more frequently with the center cell (topic).
        Cell color encodes sentiment, following the sentiment encoding in the Sentiment Scatter Plot.
    }
    \vspace*{-0.4cm}
\end{figure}\label{fig: topic_hive_first_design}

\subsubsection{Topic Hive (Initial Design)}\label{sec: topive_hive_first}
Encoding a selected topic and its highly co-occurring topics as a hive (\autoref{fig: topic_hive_first_design}),
we utilize a hive metaphor to represent ``information cocoons'' or ``echo chamber'', the notion that we are more willing to accept new information that supports our beliefs, and this, in turn, can isolate some of us in a separate informational reality.
Note that it is redesigned in the second design cycle. 
Below, we describe the first design of the Topic Hive and leave the discussion of its problems to the second design cycle.

The Topic Hive shows a deeper level of detail by visualizing the context of a selected topic.
The center cell in the hive represents the topic to be evaluated. 
The surrounding cells represent the most frequently co-occurring topics with the selected topic.
The distance of surrounding cells to the center cell encodes the frequency of co-occurrence: closer cells (topics) co-occur more frequently with the center cell (topic).
Under the hood, we sort the co-occurring topics by frequency, and bin them into ``levels''. 
Each level is a ``ring'' of hexagons that surrounds the center hexagon; i.e., the first level contains six cells, the second level contains twelve cells, and so on. 
This way, a ``higher level'' cell appears to be further away from the center cell, indicating a lower co-occurrence frequency.
We further encode the sentiment of each topic with the same color encoding as it appears in the Sentiment Scatter Plot.

\vspace*{-0.2cm}
\subsection{System Design}
NOVA's interface design stems from the interactive presentation narrative structure~\cite{segel2010narrative}.
In the initial design of NOVA, the system is divided into four stages: \textit{Topic Selection}, \textit{Belief Elicitation}, \textit{Outlet Comparison}, and \textit{Article Review}.
This multi-stage design intends to provide a loose structure for our users in their assessment of news outlets.
To support user engagement and ease of use, while users focus on hypothesis generation and verification, the system keeps track of the decisions and interactions that users make and carries them over to later stages to lessen the cognitive burden. 
We incorporate modals, user guides (tutorials), and annotations to help direct user attention and to provide context for what they are assessing. 
The system, in the first three stages, assists users to externalize their beliefs and find discrepancies between the data.
The final stage facilitates them to assess whether the discrepancies stem from their personal bias. 
Below, we briefly describe the high-level task for each stage and how a user would complete each task.

\begin{figure}[h]
    \centering
    \includegraphics[keepaspectratio, width=0.95\columnwidth]{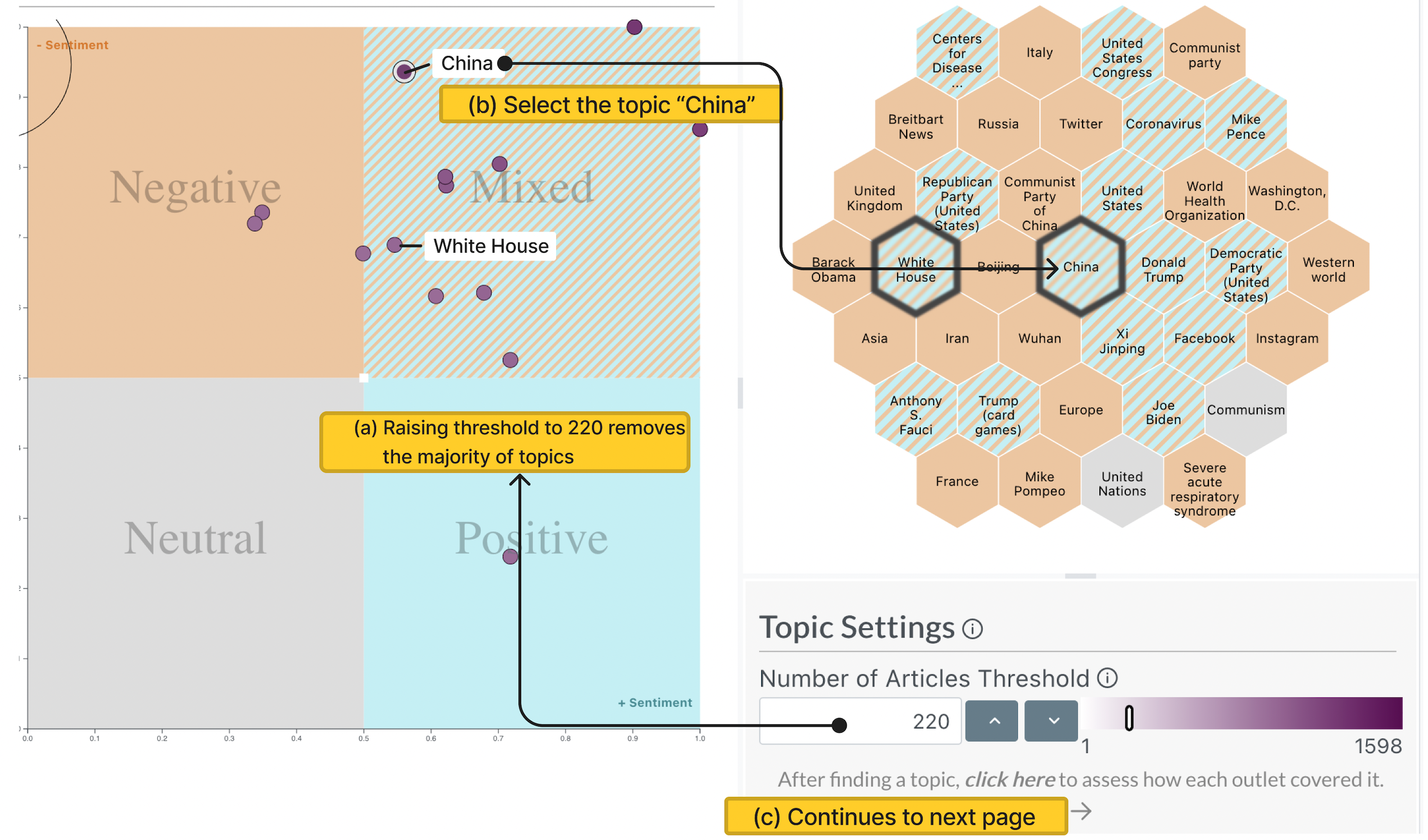}
    \caption{Topic Selection stage, users use the Sentiment Scatter Plot and Topic Hive together to discover interesting topics. An Article Threshold can be used to filter topics by the number of associated articles. Selecting a topic from the Sentiment Scatter Plot will trigger the corresponding Topic Hive. Users can choose a pair of co-occurring topics for further investigation in the next stage.}
    \label{fig: initial-overview}
\end{figure}

\subsubsection{Topic Selection}
The first stage, Topic Selection, assists users in making sense of the dataset and selecting a topic of interest.
The stage combines Sentiment Scatter Plot and Topic Hive (\autoref{fig: initial-overview}) to provide an overview of the dataset.
Users can adjust the \textit{Article Threshold} to filter out topics that are not well covered.
Clicking on a topic in the Sentiment Scatter Plot will trigger the corresponding Topic Hive to show. 
The design consideration behind this initial version is to show the sentiment discrepancies among topics, which indicates the existence of media bias. 
For example, \textit{China} appears to be significantly more negative than \textit{White House}.
This discrepancy may indicate that the media is biased against China, incentivizing users to investigate further.
In addition, the Topic Hive supports users to choose a second topic along with the center topic before navigating to the next stage.

\subsubsection{Belief Elicitation}
\begin{figure}[t]
    \centering
    \includegraphics[keepaspectratio, width=0.95\columnwidth]{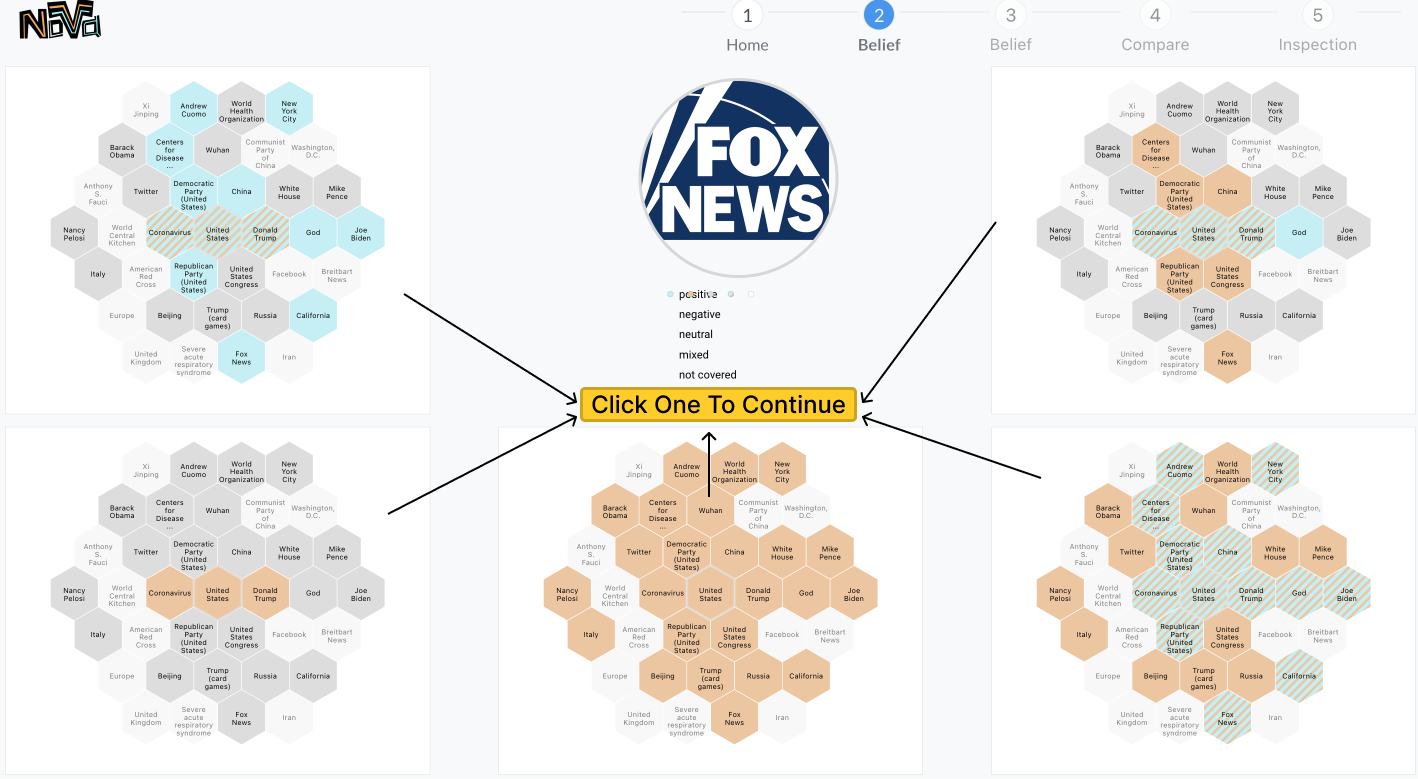}
    \caption{Belief Elicitation stage, users are prompted to select one from five randomly generated hives that best represent their belief. This process is repeated to two randomly selected outlets. The decisions made by the users are used for extrapolation in the Outlet Comparison stage. }
    \label{fig: initial-belief}
    \vspace*{-0.4cm}
\end{figure}

In this stage, the system prompts the user to answer how two randomly selected outlets covered the selected topic.
We provide five different versions of hives for the user, among which four of them are randomly generated and one of them is generated from the data (\autoref{fig: initial-belief}).
The user answers by clicking the hive that is closest to his/her belief.
After two such questions on two different outlets, we take the user input and extrapolate the answers to all six outlets.
The extrapolated results are treated as the user's belief of the coverage of the topic by the six outlets.
We decided not to ask the users to elicit beliefs on all six outlets to avoid overwhelming them. This inevitably introduced a trade-off between cognitive load and belief representation accuracy. As is shown in the pilot study, this was not a good design decision. More details are discussed in~\autoref{sec: pilot_study}.
The results are shown in the next stage, \textit{Outlet Comparison}.

\vspace*{0.1cm}
\subsubsection{Outlet Comparison}
\begin{figure}[h]
    \centering
    \includegraphics[keepaspectratio, width=0.95\columnwidth]{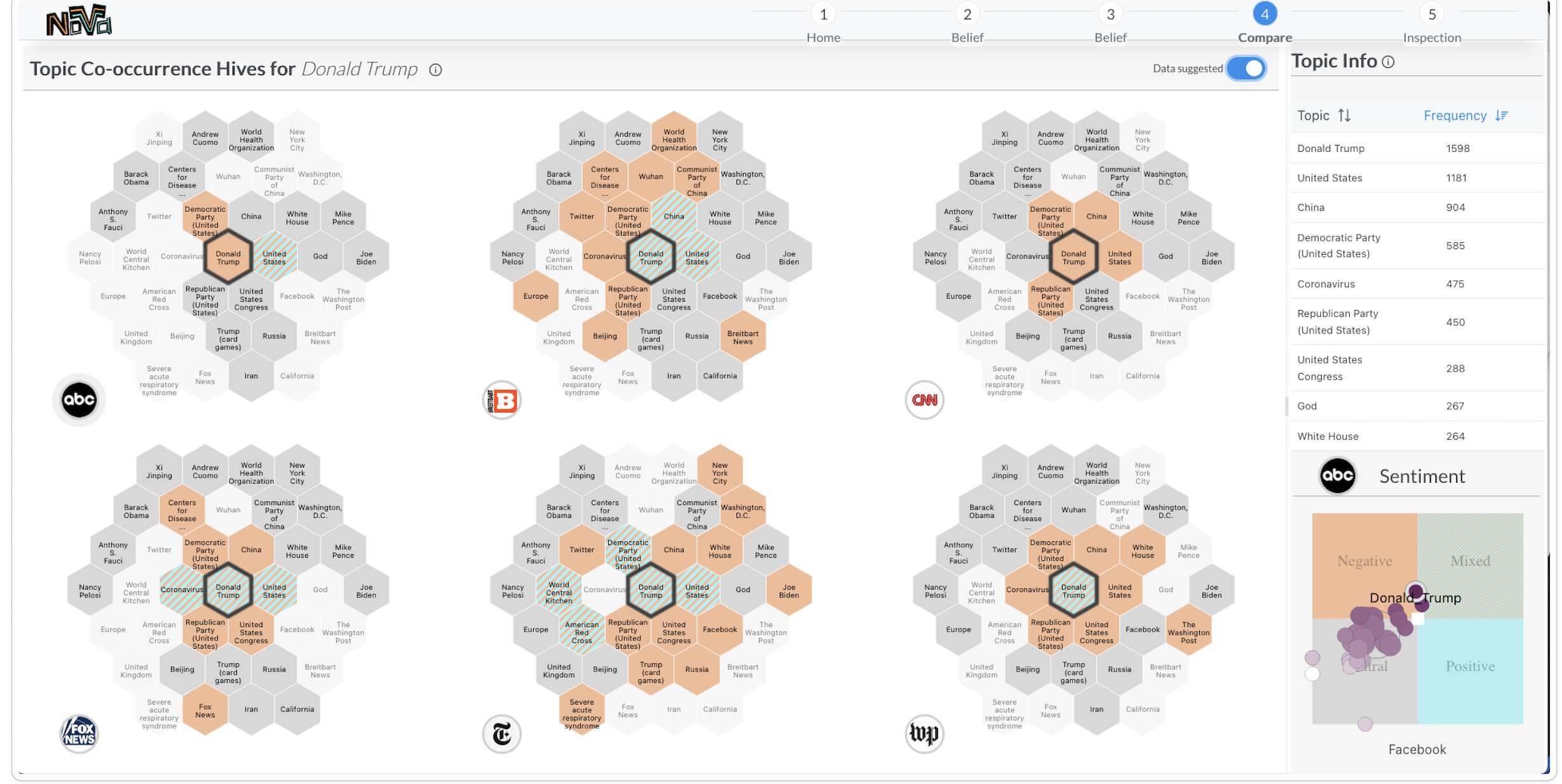}
    \caption{Outlet Comparison stage, the hives of six outlets are shown in juxtaposition for visual difference and pattern finding. We further provide a Topic Info Table to show the frequencies of the shown topics. We also show a Sentiment Scatter Plot for an outlet once the user selects a topic in the outlet. Users can select any center cell, with an optional surrounding cell, and then proceed to the next stage.}
    \label{fig: initial-compare}
    \vspace*{-0.4cm}
\end{figure}

This stage shows six hives side-by-side (\autoref{fig: initial-compare}), representing how each outlet has covered the topics within.
The comparison of the outlets supports users in generating hypotheses on media bias.
To support an easier comparison, we want the topic cells in each hive to be aligned. 
Therefore, the system first selects a fixed number of ``candidate topics'' to be encoded as cells. This selection is done by merging all the frequently co-occurring topics among the six hives and sorting them according to combined frequency. Then the sorted rankings are used to encode the positions. This ensures a fixed position across all six hives. 
Consequently, some topics might not be mentioned by an outlet at all. We use semi-transparent cells to encode missing topics. For example, in the Topic Hive of ABC News (the top-left hive in~ \autoref{fig: initial-compare}), a large number of missing topics (semi-transparent cells) can be observed in the lower-left corner.
Since the hives are generated from the user's belief, the hives should mostly fit the user's expectation, and the outliers can be easily spotted.
If the user finds any strange results or anything that contradicts what they believe, they can select the hive and go to the next stage and verify their hypothesis.

\vspace*{0.1cm}
\subsubsection{Article Review}
\begin{figure}[h]
    \centering
    \includegraphics[keepaspectratio, width=0.95\columnwidth]{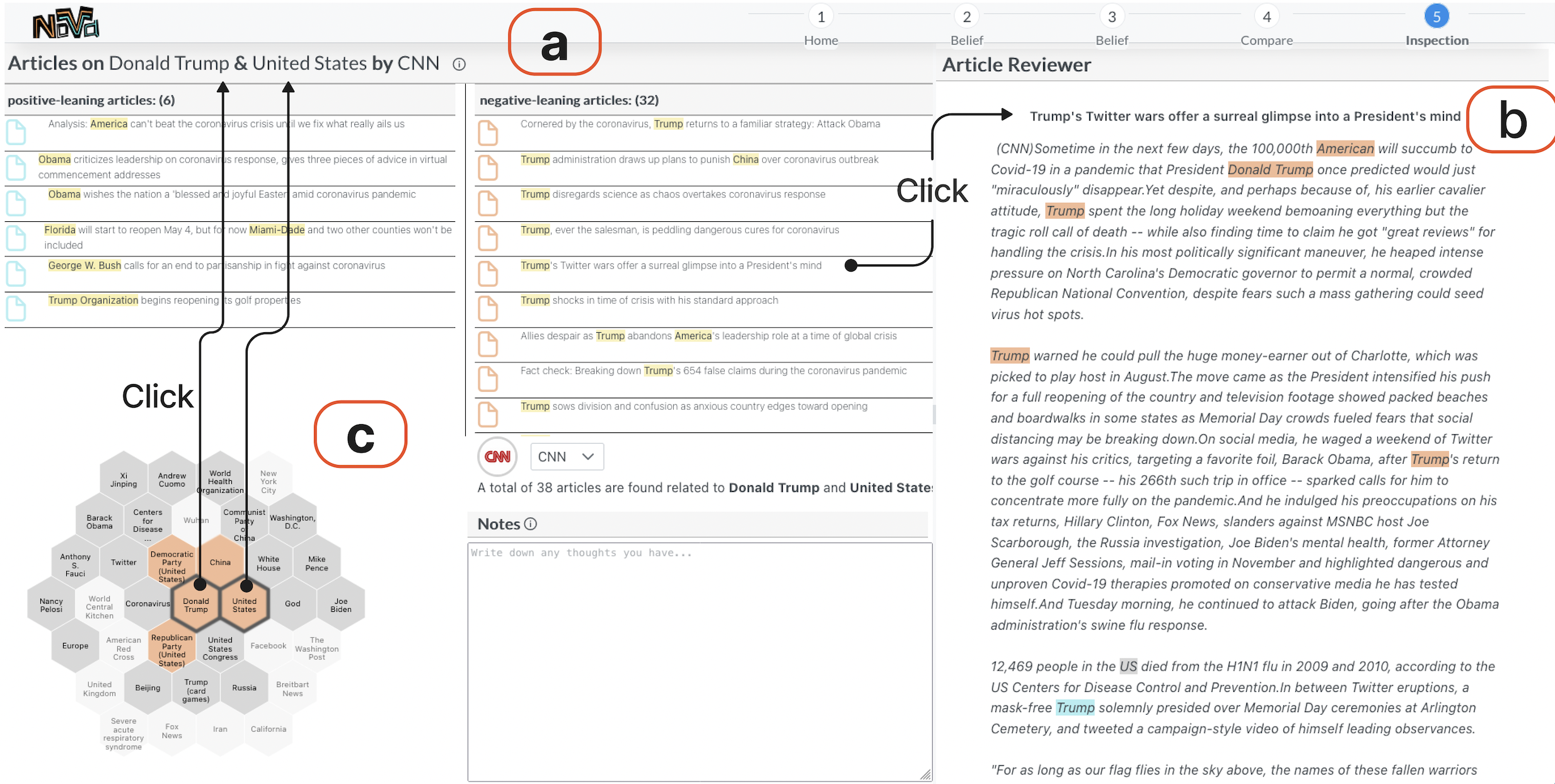}
    \caption{Article Review stage, users can choose articles in (a) and read the article in (b) to find evidence to support or disprove their hypothesis. The selected hive from the previous stage is shown in (c) as context. Clicking on a hive cell changes the selected topic.}
    \label{fig: initial-inspection}
    \vspace*{-0.5cm}
\end{figure}

In the last stage, Article Review, users can review associated articles to find evidence to support or disprove their hypothesis.
In this stage, we present \textit{Article View} (\autoref{fig: initial-inspection}-a), which depicts to users, the articles associated with the initially selected topic from Stage One and a conflicted topic selected from Stage Two. 
The upper half of Article View lists the article's headlines, classified as positive or negative with keywords extracted as described in~\autoref{sec: el}. 
We place the list of positive and negative articles in juxtaposition with each other to make it easier to compare.
Selecting a headline will load the~\textit{Article Reviewer Panel} (\autoref{fig: initial-inspection}-b), showing the full content of the selected article. 
We further highlight the mentioned entities in the color of their sentiments in the sentences.
Finally, users can record their findings in the notes panel.
We show the selected hive from the previous stage as context, which in addition can be used to change the selected topic and corresponding articles.
Evidence-finding in the last stage concludes one round of assessment. 
Users can go back to the previous stage to investigate another topic by repeating the process.

\subsection{Pilot Study}\label{sec: pilot_study}
We performed a pilot study to evaluate the usability and usefulness of the first design of NOVA.
We report our qualitative analysis results based on recorded videos and survey answers, which motivated us to conduct a second design cycle.

\subsubsection{Participants}
In our study, we recruited 12 people\footnote{The demographic breakdown of the participants was:  75\% between ages 20--29, 17\% between 30-39, and 8\% above 60, 6 females and 6 males.} to use NOVA.
Before each session, each participant filled out a survey of their U.S. news consumption as well as provided demographic information.
From the initial survey, 75\% of our participants actively seek out news content, with a majority (66.7\%) reading the news more than once a day.
Most of our participants access their news via social media (75\%) and/or an online subscription (50\%).
All our participants had read an article or seen an article headline within 2 weeks of taking the study from at least one of the 6 media outlets we gathered articles from. 
After filling out the initial survey and tutorial on NOVA, they would then be directed to our application in a Google Chrome browser.

\subsubsection{Procedure}
Participants were given 2 tasks. Task 1 requires participants to find a reported topic of interest using NOVA and write down their expectations for how it was covered and the rationale. Task 2 requires participants to assess their expectations using NOVA, by comparing differences in sentiment between their perception and the actual reporting while writing down any insights. 
The goal of Task 1 is to evaluate the effectiveness of NOVA in capturing users' beliefs on media outlet coverage for a topic, while the goal of Task 2 is to ascertain if NOVA is effective in facilitating the assessment of their beliefs toward media outlets. 
We performed thematic coding with survey results and categorized user motivations to derive what motivates user interactions in NOVA and what aspects can be further considered when designing visualization systems for personal belief assessment.

\subsubsection{Findings}
The pilot study revealed several insights into the initial design.
Overall, the participants were able to use NOVA to assess their beliefs about mainstream media coverage,
but some design flaws still exist and need to be addressed.
Below, we summarize the findings from the pilot study.
\begin{itemize} 
    \item \textit{F1: The structure works. } The multi-stage structure works well for simplifying the process.
Although navigating through multiple pages is not a familiar design, participants reported that the stage division is natural, and they always know what to do at each stage.
Furthermore, the system could support the generation of questions.
The majority of participants formulated questions about media reporting or questioned why their prior expectations were misaligned.
Half were able to either validate their expectation of a media outlet or reveal they were biased.
This shows that dividing a complex task like self-assessment of personal belief into multiple sub-tasks on multiple pages is the right design direction.
    \item \textit{F2: User engagement is low.}
The study also reveals some design flaws as we observe a low user engagement.
Some participants reported that they were not familiar with all outlets.
This unfamiliarity with some of the presented outlets would result in difficulties in answering the questions.
Participants often have to resort to random guesses to pick a desired hive, discouraging them from further usage.
Furthermore, in outlet comparison, they tend to only compare outlets they are familiar with and ignore the rest.
Presenting all six outlets side-by-side became redundant and overwhelming.
Also, the amount of visualizations on each page appears to be overwhelming. 
Users often fixate on an insignificant visual pattern and forget the task for the stage. 
All these factors combined contributed to low user engagement, consequently preventing successful belief elicitation and updating. 
    \item \textit{F3: Topic Hive is ill-designed. }
Finally, the topic hive effectively shows the frequently co-occurring topics and their sentiments,
but the positions are not intuitive to interpret and participants often ignore the frequency order.
In some cases, not aware of the frequency differences, participants would select a peripheral topic, and proceed to the next stage, only to find that the topic is insignificantly discussed by the outlet.
Also, the participants often intuitively associate topics close to each other as related, indicating that the position encoding needs to be redesigned. 
Finally, the subtle position encoding changes between the Topic Selection stage and the Outlet Comparison stage are confusing for most participants.
A key point of confusion was the semi-transparent cells.
Although semi-transparency is often used to denote missing data many users were confused in interpreting its meaning.
We find this is likely due to semi-transparent cells first being introduced on the Outlet Comparison stage and not on the Topic Selection stage.
Several participants wanted to go back to the Topic Selection stage and check if their understanding of the visual encoding was correct, disrupting the belief assessment process. 
We decided to redesign Topic Hive in our next design cycle.
\end{itemize}

\vspace*{-0.1cm}
\section{NOVA: The Second Design}
\noindent Based on the findings from the pilot study, we went through a second design cycle to improve the design.
We first introduce our design requirements derived from related works and the pilot study, and then we describe the second design in terms of changes from the first design, and how it addresses the design requirements.

\vspace*{-0.35cm}
\subsection{Design Considerations}
The primary objective of NOVA now became to enable people to assess their personal beliefs.
We combine the sense-making model proposed by Lee et al.~\cite{lee2015people} and structure our belief elicitation design under the VIBE design space~\cite{mahajan2022vibe}. 
We target users that have abundant news-consuming experience. 
Regular news consumers do not necessarily have high visual literacy or background, and this limits our design choices to be simple. 
To support the assessment of users' personal bias toward media outlets, we design a workflow in which users first externalize their belief through a visualization, then contrast the visualization to the data-driven one, and finally find evidence to explain the differences. 
To support this process, we outline the following design considerations:

\begin{itemize}
  \setlength\itemsep{0em}
  \item \textit{DC1: Support sensemaking.} The personal belief evaluation starts with the sensemaking of what has been covered and how, but news media coverage can be diverse and overwhelming to process. Thus, NOVA must ensure all visual content is easy to interpret and the public has adequate context. The design must also refrain from overwhelming the public with unnecessary details, as they are non--experts and have free--choice in participation~\cite{peck2019data}.
  \item \textit{DC2: Externalize Personal Belief.} The first step of personal belief assessment is to externalize the belief. NOVA should facilitate users in externalizing their personal beliefs and visually represent this externalization.
  The visualization should be intuitive and decipherable enough for users to understand and adjust.
  \item \textit{DC3: Compare Personal Belief and Data.} The second step of personal belief assessment is to compare the belief with the data. We recognize that externalizing beliefs as numerical values is limited in the context of news media coverage, thus we leverage the visual comparison of two visualizations as feedback. This extends \textit{DC2} in that the belief-driven visualization should not only be easy to understand and adjust but also easy to identify differences. The comparison of personal belief and data naturally leads to belief evaluation and updating.
  \item \textit{DC4: Evaluate Personal Belief through evidence finding.} Even though the user's personal belief is compared against the data, we do not assume that the data version always reflects ground truth. Therefore, we provide a way for users to navigate themselves in the data and find evidence that attributes to the differences to support or disprove their beliefs.
\end{itemize}

\noindent These design requirements are derived from existing works and the pilot study. Studies from~\cite{peck2019data,  ma2019decoding, dasu2020sea, stokes2022striking} provide guidelines for improving visualization novices' sensemaking process (\textit{DC1}) and informed us of the visualization design in the redesign of Topic Hives.
We took inspiration from Segel et al.~\cite{segel2010narrative}, that the narrative structure design of the system can help lower the cognitive load for public users, which is also supported by the pilot study (\textit{F1}).
We maintain this structure in the second design and further emphasize the importance of \textit{DC1} at each stage by adding modals that explain the purpose of the stage. An example modal is shown in~\autoref{fig: second-belief}-ii. Modals for other pages are provided in supplemental materials.

\textit{DC2} is partially supported in the initial design, and from the pilot study, we identified how the design can be further improved.
First, the externalization in the initial design failed. Users were essentially unable to select a desirable Topic Hive due to interpretability issues with the data encoding (\textit{F3}) and the overwhelming size of the selection pool (\textit{F2}).
To improve the externalization and better support \textit{DC2}, we redesigned the Topic Hive and emphasized the construction of a personal belief hive through interaction.
Since we found in the pilot study that comparison between outlets is not desirable (\textit{F2}), we switched our focus and let users compare their personal beliefs with the data, as outlined in \textit{DC3}, a more common approach that has been proven to be successful in the literature~\cite{heyer2020pushing}.

Finally, we keep the evidence-finding functionality in the second design to support \textit{DC4}. As mentioned earlier, we do not seek to persuade users that the data generated visualization represents the ``ground-truth'', and instead suggest that they are merely ``signals'' to guide users with the assessment of their personal beliefs. This design consideration is supported in the pilot study, as we observed less questioning around the data credibility and a natural need to inspect the articles in detail in every participant.
This new set of design considerations enables us to further simplify our system design into three stages, namely \textit{Topic Selection}, \textit{Belief Elicitation}, and \textit{Article Review} while supporting an even more effective belief externalization, comparison, and assessment.

\vspace*{-0.35cm}
\subsection{Topic Hive (Redesigned)}\label{sec: topic_hive_second}
Motivated by \textit{F3} and the \textit{DC2}, we redesigned the Topic Hive to better support the externalization of personal beliefs (\textit{DC2}). 
Instead of using the position to encode co-occurrence frequency, the hive is now divided into four regions to encode sentiment type: positive, negative, neutral, and mixed.
The region-division encoding inherits from the Sentiment Scatter Plot to lower the cognitive load for users to understand the encoding (\textit{DC1}).
A topic assigned to one of the hexagons implies the overall sentiment of the outlet when covering the topic.
For example, a topic in a negative color hexagon implies that the outlet would report negatively about the topic.

Throughout multiple stages in NOVA, a topic hive can be generated through user interactions (\textit{DC2}) or automatically generated from the data.
Users construct the hive by drag-and-drop unassigned hexagons into the hive.
When automatically generated from the data, the classification of each topic is assigned by the segmentation rule specified in \textit{Sentiment Scatter Plot}.
This flexibility in the hive generation enables us to compare user-generated hive and data-generated hive, thus supporting \textit{DC3}. 
Also, by dividing the hive into regions, the differences between the user hive and the data-generated hive can be more easily recognized.
As opposed to a comparison between six outlets in the first design, this second design focuses on comparison between only two hives, further lowering the cognitive load for the users. 
In the second design of NOVA, Topic Hive is the main visualization for users to externalize their personal beliefs and motivate self-assessment.
More details about the usage of Topic Hive throughout the multi-stage system are described in the next section.

\vspace*{-0.35cm}
\subsection{Interface Design}\label{sec: interface_design}
The two main objectives mentioned in~\autoref{sec: first_design} remain the same.
To address the first objective we take advantage of narrative visualization techniques and strategies to pace out the assessment and facilitate users to have more control over their process. 
To address the second objective NOVA supports belief elicitation by facilitating users to express their beliefs about the topic and the outlet they choose and contrast their beliefs to the data.
The system is redesigned into three stages: \textit{Topic Selection}, \textit{Belief Elicitation}, and \textit{Article Review}. Additionally, users can go through multiple ``rounds'' of assessment, where in each round a different pair of topics and outlets is assessed. 
Below, we describe each stage in more detail.

\subsubsection{Stage One: Topic Selection} 
\begin{figure}[t]
    \centering
    \includegraphics[keepaspectratio, width=\columnwidth]{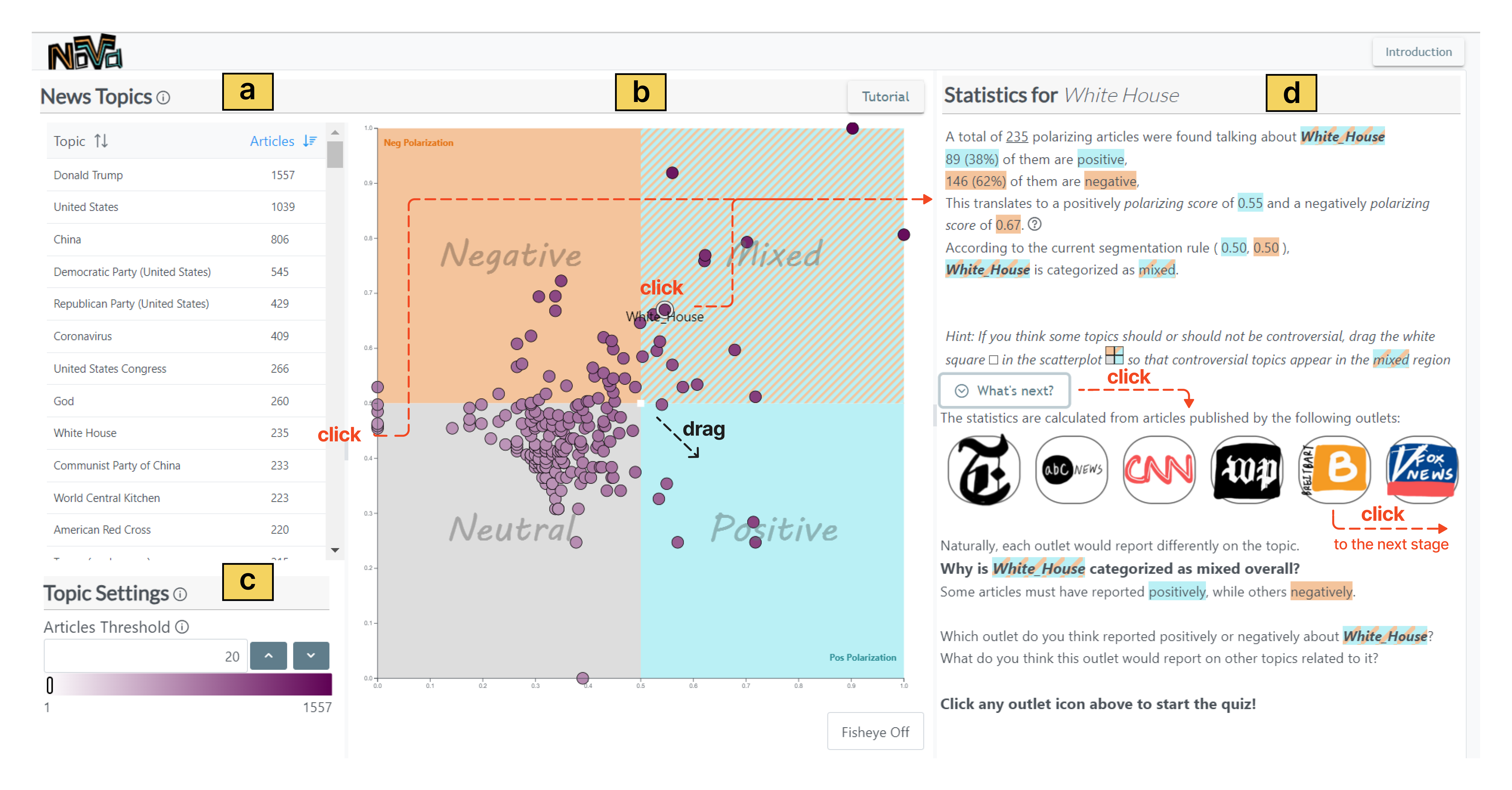}
    \caption{Topic Selection stage. (a) A table of entities and their number of mentioned articles. (b) The sentiment scatter plot shows two-dimensional sentiment for each topic. The region is divided into four categories: mixed, positive, negative, and neutral. (c) The utility panel contains a filter and a color scale legend on article frequency. (d) By choosing a topic from (a) or (b), users can see the statistics for the topic, and they can choose an outlet to inspect further.}
    \label{fig: second-overview}
    \vspace*{-0.5cm}
\end{figure}

The first stage, Topic Selection, is still focused on easing users into the system and facilitating them to get a sense of the data (\textit{DC1}) and pick a topic to delve into.
We kept the Sentiment Scatter Plot and the Article Threshold in the initial design (\autoref{fig: second-overview}-b and -c) and additionally added a Topic Table (\autoref{fig: second-overview}-a) to provide a list of topics and their corresponding article numbers.

Instead of showing the topic hive in the initial design, after selecting a topic, NOVA will now show the statistics data of the topic in a narration (\autoref{fig: second-overview}-d), explaining how the numbers of positive and negative articles related to the sentiment and how we categorize the sentiment type of topic by the score (\textit{DC1}). 
The navigation hints to the next stage are hidden by default and can be shown by clicking the ``What's Next'' button under the statistics explanations.
Once clicked, NOVA shows the six mainstream media outlets and uses prompts to motivate the user to choose one for further assessment.
Note that in the second design, NOVA directs the user to choose only one outlet instead of showing all six outlets for further assessment.
This design choice is informed by \textit{F2}, that users prefer to focus on assessing one outlet at a time. 
Once a topic and a media outlet are chosen, the assessment begins in the next stage, \textit{Belief Elicitation}.

\subsubsection{Stage Two: Belief Elicitation}\label{sec: belief_stage}
\begin{figure}[t]
\renewcommand*\thesubfigure{\roman{subfigure}} 
    \centering
    \subfloat[\centering
    ]{\includegraphics[keepaspectratio, width=\columnwidth]{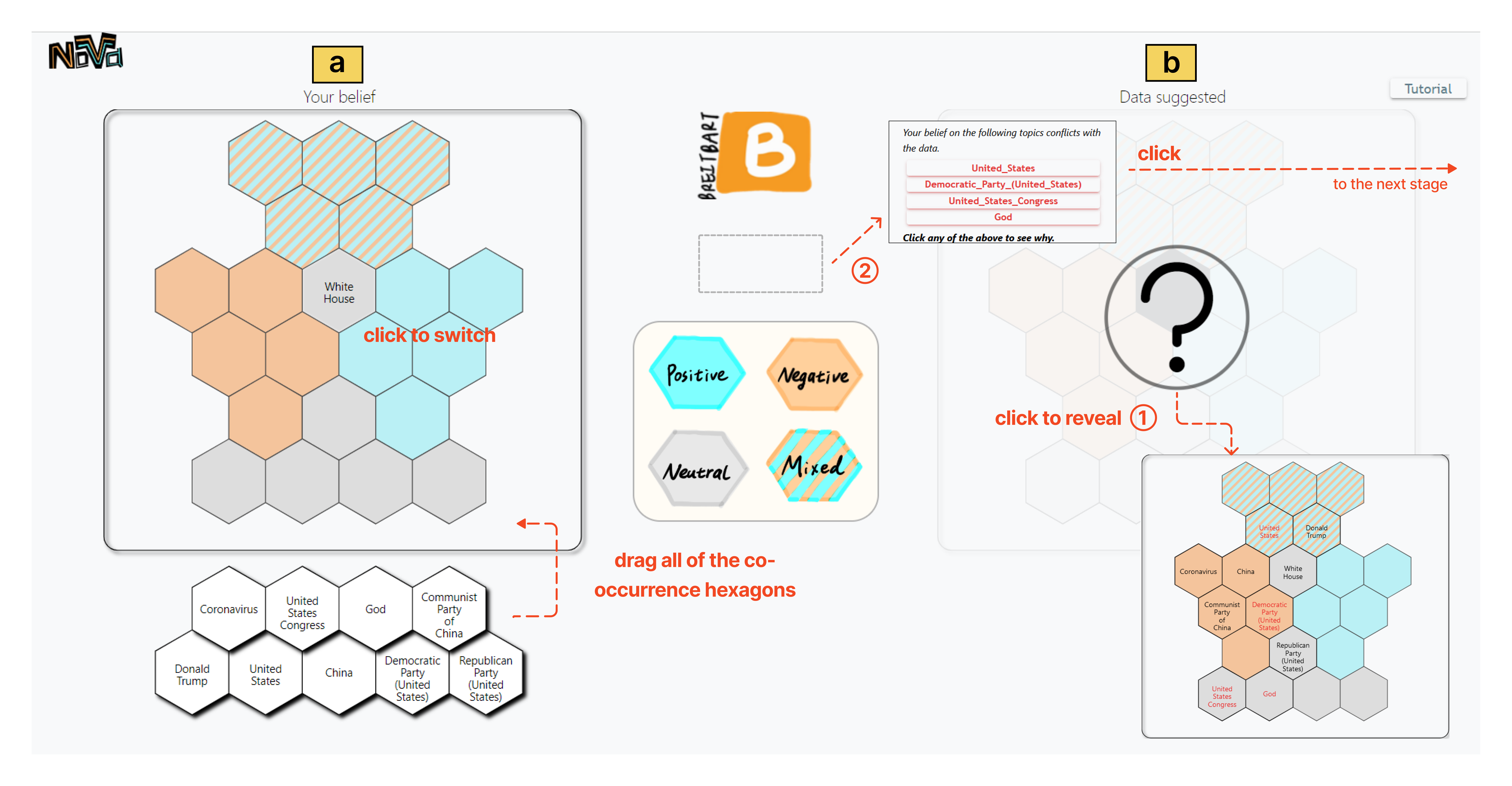}}
    \qquad
    \subfloat[\centering  \label{fig: modal-belief}]
    {\includegraphics[keepaspectratio, width=0.95\columnwidth]{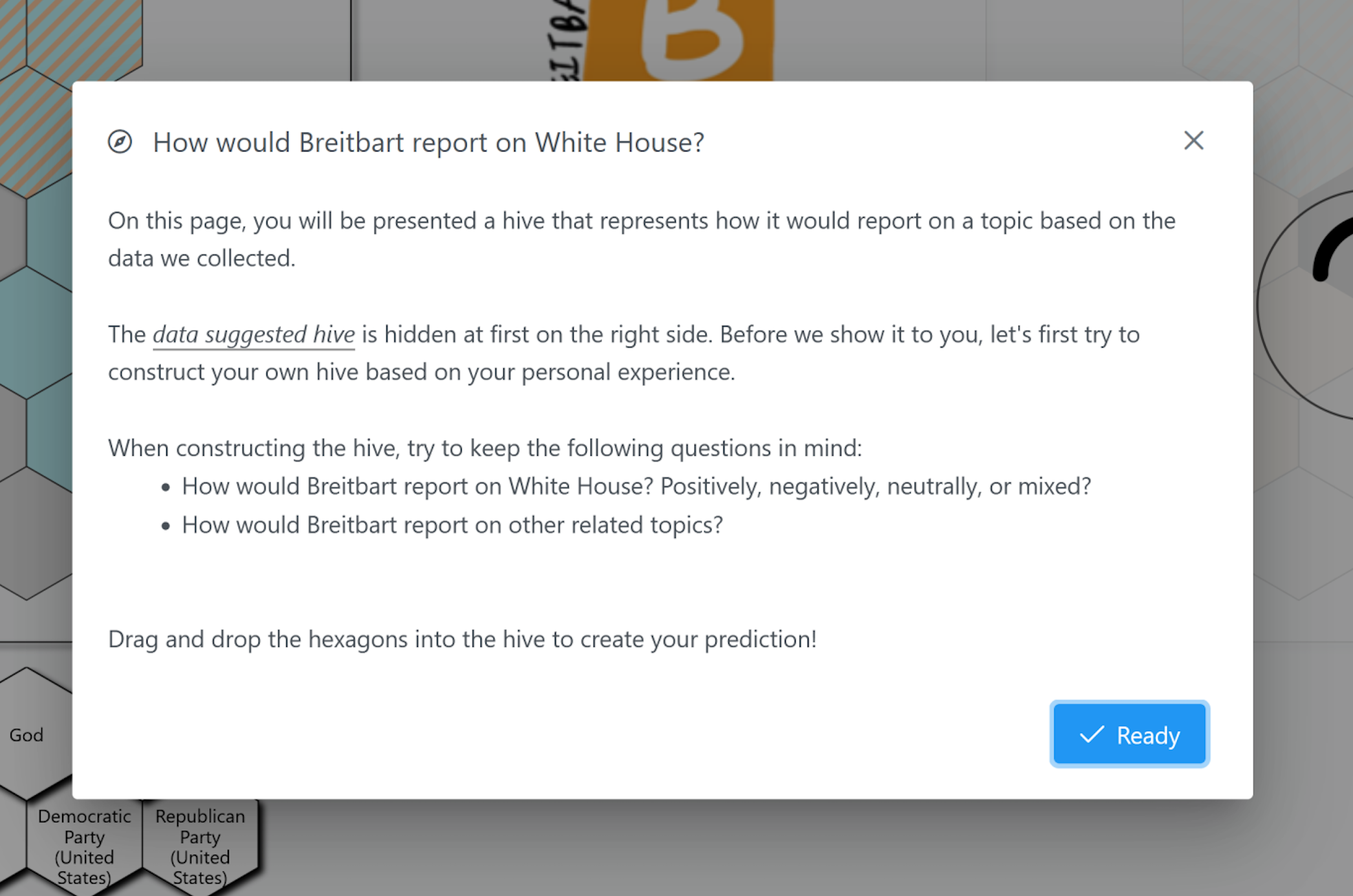}}
    \caption{
    (i):  Belief Elicitation stage. (a) To externalize a user’s belief, the user can drag the topic hexagons to the corresponding sentiment category or click the center hexagon to adjust its sentiment. (b) After clicking the question mark, NOVA reveals the data hive and highlights the discrepancies between the user’s belief and the data, motivating the user to investigate the conflicts.  
    (ii):  A modal explaining the purpose of the Belief Elicitation stage.
    }
    \label{fig: second-belief}
\vspace*{-0.5cm}
\end{figure}

The Belief Elicitation stage takes user input on perceived media outlet coverage. 
We redesigned this stage completely by changing the belief elicitation strategy, visualization, and interactions. 
From the previous stage, the user chooses the outlet which he/she is the most familiar with. 
NOVA will then generate other topics that are highly associated with the selected topic, as shown in the bottom part of~\autoref{fig: second-belief}i-a. 
Each hexagon represents a topic, and the color of the hexagon is either positive, negative, neutral, or mixed. 
In this stage, we collect the user's beliefs by instructing the user to drag each of the topics to the colored hexagon (\textit{DC2}), constructing a Topic Hive of their belief.
The visual encoding of the topic hive is deliberately designed to be loose: the five slots in the same region are equal to each other, without any relevancy or importance encoding.
The loose encoding accounts for uncertainty when people reflect on their perception of media coverage (\textit{DC1}), i.e., users tend to be uncertain about the coverage of the topics, especially those they are not familiar with. 
Enforcing a strict encoding would force the users to make hard decisions that ultimately impair the truthfulness of the constructed hive. 

After users inject their personal beliefs, we reveal the data-generated version of hives for comparison, as shown in~\autoref{fig: second-belief}i-b. Furthermore, NOVA assists users in discovering the differences by highlighting the conflicting topics in red text color (\textit{DC3}). The result is shown as a conflict topic list in the middle of two hives. Users can pick one from the list and find out the possible reasons that cause the difference in the next stage.
In the descriptions of the interface, we emphasize that the differences are ``conflicts'' rather than ``errors'', a major distinction of NOVA.
NOVA is a platform to assess personal beliefs on media outlet coverage, and the data-driven topic hive is simply a ``signal'' that directs users to possible biases. In the next stage, \textit{Article Review stage}, users can try to find evidence to support or disprove their personal belief, and then decide for themselves whether it is a bias.

\subsubsection{Stage Three: Article Review}\label{sec: article_stage}
\begin{figure}[t]
    \centering
    \includegraphics[keepaspectratio, width=\columnwidth]{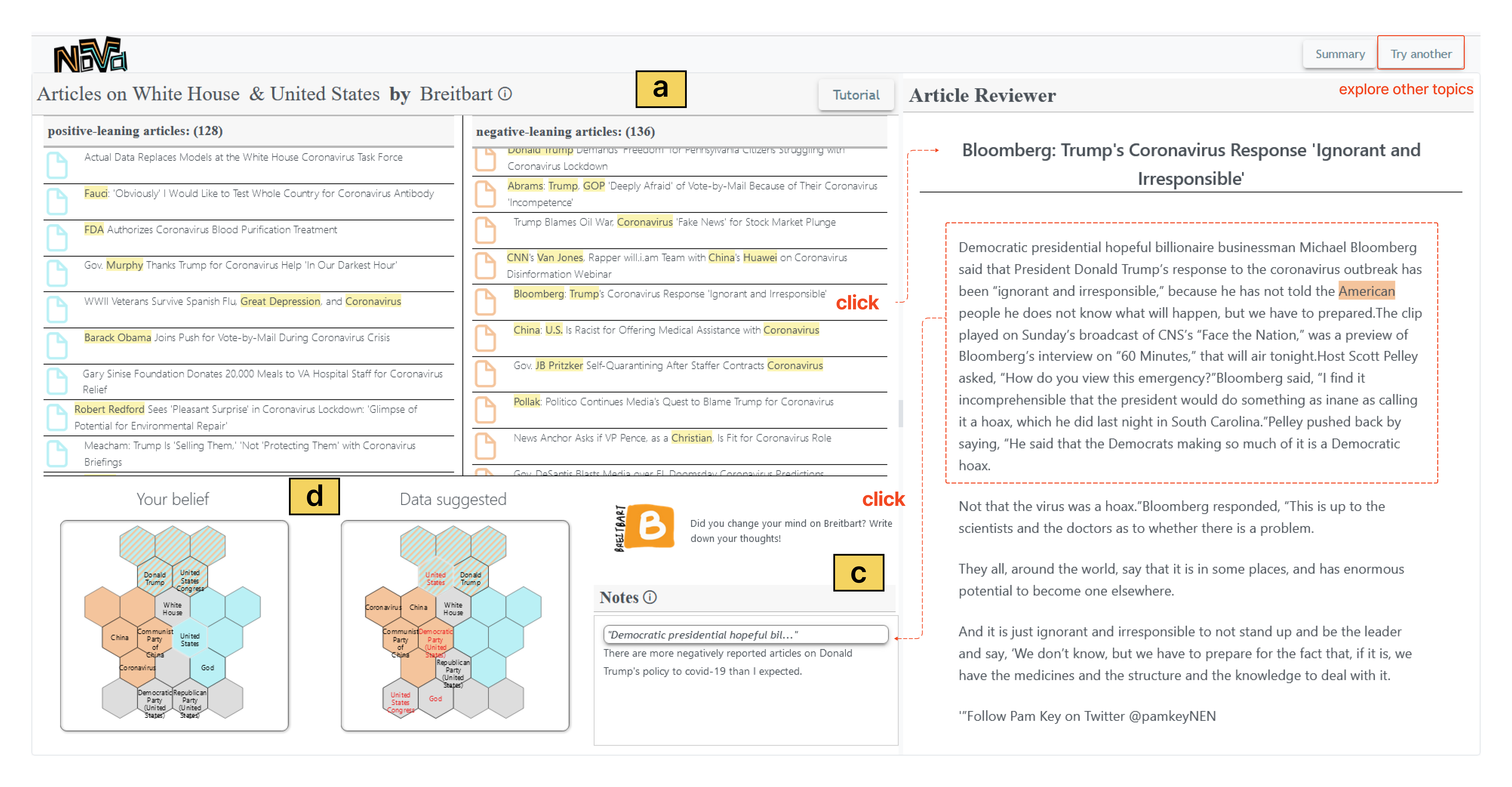}
    \caption{Article Review stage. (a) Article Panel shows positive and negative articles in two columns. (b) depicts the annotated content of a selected article. (c) Notes Panel for documenting insights. Clicking the paragraph in (b) creates a ``reference'' in the note. (d) The user and data hives are displayed. Users can click a hexagon to inspect its articles. After the exploration, users can click the ``Try another'' button on the top-right corner to explore another outlet.}
    \label{fig: second-inspection}
    \vspace*{-0.5cm}
\end{figure}

The Article Review stage remains the same as in the first design for the most part. It still focuses on belief assessment,~\textit{DC4}, where users find evidence (articles or sentences) to support or disprove their personal beliefs.
Users can still use the \textit{Article View} and \textit{Article Reviewer} to select and inspect articles in detail and record their findings in the \textit{Notes Panel} (\textit{DC1}).
An additional feature is that users can click a paragraph in the Article Reviewer and add it to the Notes Panel as a reference, as shown in \autoref{fig: second-inspection}-c.

Similar to the initial design, the \textit{Outlet Coverage} view (\autoref{fig: second-inspection}-d) also serves as a reminder of users' previous line of inquiry up to this point. 
They can use this hive to switch to a different co-occurring topic, which in turn loads different articles to browse.
At this point, users either have found evidence via the articles that help address their initial question and may feel done, or they may seek to return to Stage One and start a new line of inquiry.

The Article Review is the final stage of one iteration. 
In each iteration, users would assess their belief of how a media outlet has reported on a specific topic. We recognize the diversity of media coverage, so we encourage users to do multiple rounds of iterations.
To return to the start, they can press the ``Try another'' button or click the NOVA logo in the top-left corner of the page. 
All constructed hives and notes are saved in the summary panel (\textit{DC1}), which will be shown by clicking the ``Summary'' button, and they are free to cycle back and forth with NOVA till they are satisfied.

\vspace*{-0.2cm}
\section{Usage Scenario: Coverage of White House}
\noindent Let us walk through an example scenario of how one could use NOVA. 
We begin with our user opening up NOVA in her web browser and viewing the Topic Selection stage. After closing the modal, the stage only shows the Topic table and the Sentiment Scatter on the left-hand side, with a text prompt on the right-hand side asking the user to hover over the news topics (\autoref{fig: second-overview}). She notices a majority of topics are clustered in the lower left region labeled as neutral. However, she is interested in topics that are well covered and raises the article number threshold to 220. 

After the adjustment, she recognizes several topics, which she personally associates to be either negative or polarizing from her memory of that time. She clicks the topics of \textit{Donald Trump} and the \textit{White House} to see the detailed statistics. The narratives in~\autoref{fig: second-overview}-d help her better understand how the sentiments are generated, thus calibrating her expectations. 
To her surprise, she finds that the ``White House'' received a polarizing ratio of positive (89) and negative (146) articles.
From personal experience, she hypothesizes that ``Breitbart'' must have reported excessively more negatively. She decides to investigate deeper on Breitbart's coverage of the White House.
She clicks ``Breitbart'' to go to the next stage.

In the next stage, the system first presents modals and tutorials on the topic hive and interactions to construct a hive (\autoref{fig: second-belief}). Once finished, she is prompted to construct her own hive and contrast it with the data-generated hive. 
As an avid reader of the news, she recalls some prior articles to determine the sentiment of the outlet. 
She proceeds to drag and drop the hexagons into the hive. 
Once she is done, she clicks the question mark to reveal the data-generated hive.
The system shows 4 discrepancies, suggesting that Breitbart has reported these topics differently than she believed. 
These discrepancies are highlighted in red, and she is prompted to select a discrepancy to investigate (\autoref{fig: second-belief}-b). 
She selects ``United States'', as this is her most familiar topic, and she suspects that she would be wrong about it. 
Upon arriving in the Article Reviewer stage, she notices that the ratio of positive-leaning articles (114) and negative-leaning articles (158) is in contradiction to her belief that Breitbart has been reporting negatively. So she starts to review the positive-leaning articles. Upon close inspection, she finds that the positivity is mainly attributed to the positive attitude about the COVID-19 situation, with many politicians expressing faith in the country's health system and belief that COVID-19 is not as serious as it is claimed to be. 
Although she did not consider this in previous stages, this is still in accordance with her belief. 
However, following this logic, she should find dominantly more positive articles from Breitbart, but she finds the negative articles to be significant as well.
To her surprise, Breitbart not only supported Trump's White House but also criticized the COVID policy. In fact, Breitbart reported more negative articles than positive ones, signaling an attitude shift as the COVID-19 situation continued to progress. She records the articles, adds comments in the notes panel, and finishes this iteration of the personal belief assessment. 

During the process, the system constantly provides signals for her to explore deeper. 
In the Topic Selection stage, she can make sense of the dataset and sentiment calculations through interactions with the scatterplot, filters, and narratives. 
These signals, together with her personal experience, guide her to find an interesting topic and outlet to assess.
In the Belief Elicitation stage, she constructs the hive by recalling her news-consuming experience and then contrasts it to the data-driven one. The discrepancies serve as signals for her to investigate possible bias in her belief.
Finally, through the exploration of targeted articles, she is able to calibrate and even correct her bias. 
The system serves as a platform in which she can easily navigate through complex media coverage data and self-assess her beliefs on media outlets.

\vspace*{-0.2cm}
\section{User Study}
\noindent A goal of NOVA is to support individuals in the assessment of their personal beliefs on news media coverage.
The assessment of one's personal beliefs is a contentious task and presents challenges in accurately evaluating.
To evaluate the effectiveness of NOVA's second design, we designed a study with a larger group of participants.
The study results reveal participant's preferences when assessing their personal beliefs through an interactive data-driven visualization and offer insights into the relationship between design elements and personal belief elicitation.

\vspace*{-0.2cm}
\subsection{Study Design}
\begin{figure}[t]
    \centering
    \subfloat[\centering Age group distribution\label{fig: age_group}]{{\includegraphics[width=0.43\columnwidth]{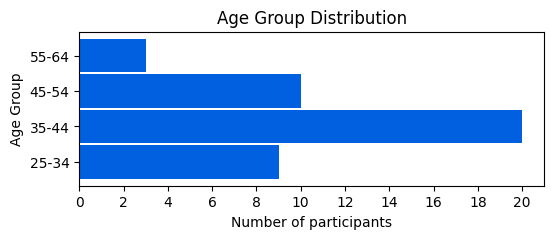} }}%
    \qquad
    \subfloat[\centering Education level\label{fig: education_level}]{{\includegraphics[width=0.43\columnwidth]{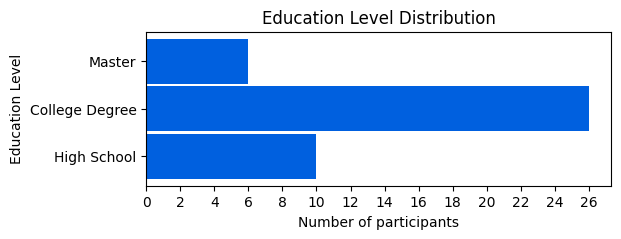} }}%
    \qquad
    \subfloat[\centering News reading frequency\label{fig: news_reading_frequency}]{{\includegraphics[width=0.43\columnwidth]{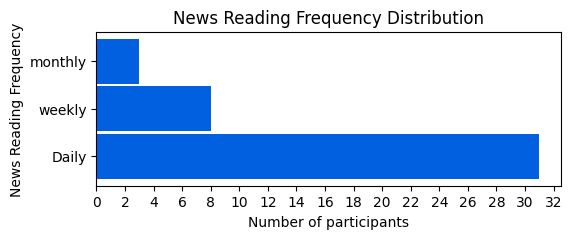} }}%
    \qquad
    \subfloat[\centering Political affiliation\label{fig: political_affiliation}]{{\includegraphics[width=0.43\columnwidth]{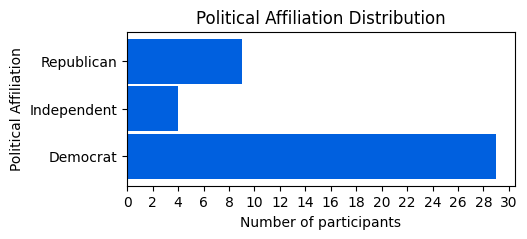} }}%
    \captionsetup{belowskip=-14pt,aboveskip=3pt}
    \caption{
        Demographic breakdown of the participants.
    }%
    \label{fig: demographics}%
\end{figure}

We recruited 42 participants on Amazon's Mechanical Turk (MTurk), with obvious spammers removed.
Additionally, we required participants to be from the United States to ensure their familiarity with US-based outlets, as our dataset consists of articles primarily produced by US mainstream media.
To ensure high-quality results, we also require participants to have a  human intelligence task (HIT) approval rate of at least 95\%.
All 42 participants receive a \$5 compensation.
For demographics, we collected their age, highest level of education, news reading frequency, and political affiliation, as shown in~\autoref{fig: demographics}.
We observe a wide range of age groups (25--54) and education levels (from high school diploma to Master's Degree).
Among the 42 participants, 9 (21.4\%) are between 25--34, 20 (47.6\%) are between 35--44, 10 (23.8\%) are between 45--54, and 3 (7.1\%) are between 55--64.
For education levels, 10 (23.8\%) have a high school diploma or GED, 26 (61.9\%) have a college degree, and 6 (14.3\%) have a Master's degree.
The political affiliation distribution appears to be heavily skewed: 29 (72.5\%) of the participants self-reported to be Democrats, 9 (22.5\%) self-reported to be Republicans, and 4 (5\%) self-reported to be Independents.
In addition, the majority of the participants (31, 73.8\%) read news daily, others read several times a week (8, 19\%) or several times a month (3, 7.1\%).
As there are no restrictions on age, education levels, and political affiliation during recruitment, the demographic distribution reflects the willingness to self-assess for different groups of people.

\vspace*{-0.3cm}
\subsection{Procedure}
We hosted the system online\footnote{https://samlee-dedeboy.github.io/Nova/} so the user study can be conducted fully remotely.
To familiarize the participants with the system, we provide a slide-show tutorial before showing participants the system.
The tutorial covers the context, the goal of the system, the dataset, visual encoding, and the system interface.
After the tutorial, the participants automatically enter the system and start free exploration.
To ensure that the participants used the system, we put the survey link in the summary pop-up in the last stage of the system (Article Reviewer).
Participants must at least navigate once to the last stage, find the survey link, and complete the survey to receive compensation.
The survey includes questions about the participants' demographics, the usability of the system, and open-ended questions about their takeaways from the system. A full breakdown of the survey questions for usability and participant takeaways is shown in supplemental materials.
We report our findings in the following sections.

\vspace*{-0.3cm}
\section{Results}
\noindent The feedback we got from participants is mostly positive: although self-assessing personal belief is a cognitively demanding process, the majority of participants find the system intuitive and user-friendly. 
Many participants gained meaningful takeaways after using our system. 
Below, we present the results in terms of \textit{Usability} and \textit{Participant Takeaways}.

\vspace*{-0.2cm}
\subsection{Usability}
We evaluate usability from three aspects: \textit{Visualization Clarity}, \textit{Interaction Intuitiveness}, and \textit{Sense-making Support}.
For each aspect, participants are prompted with a ``Yes/No/Maybe'' question, then provide textual feedback in the survey.

\subsubsection{Visualization Clarity} We asked \textit{Is the meaning of the sentiment scatter plot/Topic Hive clear to you?} (as two separate questions).
All participants found the meaning of both Sentiment Scatter Plot to be clear, and only one participant reported ``Maybe'' for the Topic Hive. 
The reported confusion around Topic Hive is from the mismatch between the topic and the associated articles.
This confusion could be a result of the use of entities to represent topics.
The main topic of an article may not be the entity that we use in the hive.
However, we consider this as a minor issue since only one participant reported confusion.
\subsubsection{Interaction Intuitiveness} We asked \textit{``Is the system functionalities/interactions (selection, filter, navigation to the next stage)'' clear to you?}.
35 (83.3\%) participants found the system functionalities and interactions clear, and 7 (16.7\%) participants reported ``Maybe''.
In the written responses, two participants reported that they had problems with the navigation, but did not give specific details.
Five participants reported that learning to use the system requires non-trivial effort.
In particular, two participants acknowledged that it is not the fault of the system.
The task of assessing personal beliefs is inherently a complex task, with many sub-tasks involved.
To support such a task, the system inevitably requires some learning effort.
However, as a majority of our participants were able to navigate and interact effectively with the system we find the learning effort NOVA imposed is acceptable.
\subsubsection{Sensemaking Support} 
We asked \textit{``During the process, have you ever found yourself not knowing what to do next?''}.
The goal of this question was to ascertain if participants either became too overwhelmed with what was presented or if the system design was uninterpretable.
The majority of the participants (32, 76.2\%) reported ``No'', 5 (11.9\%) participants reported ``Yes'' and 5 (11.9\%) reported ``Maybe''.
When reviewing the written responses, we find that the participants who reported ``Yes'' or ``Maybe'' did not have a problem with the system's sensemaking support.
Rather, they could not find the survey button, which was deliberately hidden.
The sensemaking support of the system appears successful as most participants acquired insights from the system and had many takeaways, e.g., ``\textit{The negative topics on Donald Trump reminded me of how awful he was as a president during the time of the pandemic}'', ``\textit{Seeing what different publications thought about foreign powers was certainly fascinating}''.

\vspace*{-0.2cm}
\subsection{Participant Takeaways}
To understand whether NOVA supports individuals in assessing their beliefs, we asked a series of questions in the survey to gain a deeper insight into participants' takeaways from using the system.
We report the results in the following:

\subsubsection{Rounds of Assessment}
According to the survey, 21 (50\%) participants assessed their beliefs for 3 or more rounds, 17 (40.5\%) participants assessed their beliefs for 2 rounds, and 4 (9.5\%) participants assessed their beliefs for 1 round.
This implies that the system is easy to navigate and interesting enough to attract attention.
The majority of topics being assessed are centered around COVID-19 (as our dataset consists of articles from Jan to June 2020), such as Coronavirus, Donald Trump, the White House, and China.

\subsubsection{Contradictions}
We asked in the survey \textit{``Did you find any contradiction between your belief and the data? If so, what is the contradiction?''}
We divided the responses into three categories: Overestimation of Polarization, Underestimation of Polarization, and No Contradiction.
22 (52.4\%) participants overestimated the polarization of a media outlet, 7 (16.7\%) participants underestimated the polarization of a media outlet, and 13 (31\%) participants did not find any contradictions.
This is in line with the conclusion from previous studies that the American people tend to overestimate the polarization of the media~\cite{hardy2017jittery}.
Note that the system's goal is not to convince the participants that their beliefs are wrong, but to help assess the alignment between their beliefs and the data.
This means the participants might not find any contradictions and reaffirm their beliefs, which is a valid assessment.

\subsubsection{Verification}
The survey prompts \textit{``Do you find any articles to support or disprove your belief? If so, what are the articles about? How do they support or disprove your belief?''}
Among the 42 participants, 22 (52.4\%) participants found articles to disprove their beliefs, 9 (21.4\%) participants found articles for approval, and 13 (31\%) participants did not find any articles for approval or disapproval.
Note that a participant could find articles for both approval and disapproval.
Moreover, among the 22 participants who found articles to disprove their beliefs, 5 of them stated that their beliefs did not change.
These results help support our system design choices for creating space for belief elicitation without pressuring individuals to alter their beliefs. 
By creating such a space, individuals can assess and explore these beliefs at their own pace and discretion.

\vspace*{-0.1cm}

\section{Discussion}
Belief updating is a complex process, especially when the beliefs are personal and subjective as in the case of political biases within the media.
With NOVA, the users go through this process in a step-by-step manner. 
While this ensures a gradual experience, it is still a cognitively demanding process for the user, which in turn makes the system design challenging.
In this section, we discuss our findings from the user study and the implications for future research.

\subsubsection{The three-stage design is still demanding for minorities.}
One of the design considerations of NOVA is to support the sense-making process and lower the cognitive load of the users.
While the narrative structure design was proven successful in the pilot study and the majority of the participants found the system user-friendly in the user study, some participants still find the system cognitively demanding.
For example, one participant reported that \textit{``I had a hard time overall navigating the whole system, there was a lot of information and I was overwhelmed a bit.''}
As observed in the demographics of our participants, audiences who might be interested in the topic range from high school students to senior citizens.
The three-stage design is a good start, but it requires the user to navigate back and forth between pages, which can be a burden for some users who are not familiar with the computer interfaces in general.
Moreover, since the system is heavily based on visualization and narratives, the cognitive load can be affected by the user's visual literacy and reading speed.
A more carefully designed interface that caters to minorities is needed to appeal to a wide range of audiences.

\subsubsection{Accounting for varying backgrounds and beliefs.}
The user study shows that the participants have varying perceptions of the media outlets.
Some participants overestimate the polarization of the media: \textit{``I felt that certain news agencies were more negative than I would have guessed about topics I view favorably.''};
while others underestimate the polarization: \textit{``More neutrality than I had anticipated''}.
Some even have directly conflicting perceptions over the same media outlet.
This confirms the decision to design NOVA as a platform for self-assessment, rather than a system that persuades the users to change their beliefs.
The belief elicitation techniques play a crucial role in the self-assessment design of NOVA. 
Still, more importantly, the system simply points out the \textit{differences} between the users' beliefs and the data, leaving the interpretation to the users.
This strategy avoids the problem of users disagreeing with the system and the distrust and frustration as a consequence.
We believe that this is a promising strategy for other systems that deal with users from varying backgrounds and beliefs on provocative topics. 

\subsubsection{Lessons Learned for Visualizing Belief Elicitation.}
The redesign of the Belief Elicitation stage, after the pilot study, showed promising results. We discuss lessons learned from this redesign experience. The first design of the Belief Elicitation stage prompted users with two rounds of surveys; where each round displayed five randomly generated hives and requested users to choose one that best fit their beliefs. 
Then the system infers from these two user decisions to visually ``guess'' the user's beliefs. Although statistically reasonable, a few factors could hinder the effectiveness of the system. 

First is the firmness of the belief. In the first design, the user was surveyed with two randomly selected outlets that they were not necessarily familiar with. This unfamiliarity casts self-doubt on their decision: even if they made a choice, the choice would not reflect their true ``belief''. In such cases, the belief elicitation fails due to the users not firmly believing their choices. Even if their choices are challenged by the data, users can dismiss the challenge as a lack of knowledge or context, and self-reflections are not triggered. The solution in the second design is to let users pick only the outlet they're interested in for assessment. 

The second change was to emphasize user engagement in the design. As opposed to letting users click to choose a desired hive, the second design enables users to construct a hive from scratch. With a more engaged user experience, users are more familiar with their constructed hive and recognize when the data hive suggests a conflict, incentivizing them to investigate further. 
We recognize that the construction of a hive is more cognitively demanding than selecting one. To facilitate a user-friendly construction process, we redesigned the encoding of Topic Hive to prioritize interpretability.  We found these design improvements for facilitating user engagement to be especially impactful in motivating the general public to engage in the belief elicitation process.

Our approach for the above issues was shown to be effective in the follow-up user study: we had minimal negative feedback about the effectiveness of the constructed hive; most participants gained meaningful takeaways from the system. We contribute to the belief elicitation field by showing that firmness of belief and user engagement plays a crucial role in belief elicitation. 

\subsubsection{Self-assessment is a promising direction, but further research is needed to include real-life factors.}
From the user study, 31 out of 42 participants were able to find articles to either support or disprove their beliefs.
Among them, participants frequently expressed surprise over the articles that they found.
For example, one participant reported that \textit{``I was surprised that CNN had so many negative stories on the Democratic Party.''}.
This shows that through belief elicitation and assessment, NOVA guides users to articles that they would not normally read.
On the other hand, many participants expressed explicitly that they confirmed beliefs after reviewing the articles:
\textit{
``I saw many articles for both supporting and opposing my views on a topic, but nothing to change my beliefs one way or another on any given topic.''
}
This highlights the challenge in personal and subjective belief updating, that simply showing people the data or surprising them is not enough to change their beliefs. 
In our user study, one participant reported that \textit{``I think I have learned a lot and just one session of reading articles probably won't change some of my deep-held opinions.''}
This shows that personal beliefs are deeply rooted in one's real-life experience and social circle. 
A news article dataset can only account for online information exposure, which is only one of the many factors that shape one's beliefs.
We believe that self-assessment is a promising direction, and further research is needed to understand other factors that can not be captured by online news articles and design systems that can account for real-life factors.

\vspace*{-0.1cm}
\section{Limitations and Future Work}
\noindent NOVA enables general audiences to freely assess one’s personal beliefs toward media outlets. 
It is designed to be accessible and easy to use.
Feedback from the user study shows that NOVA is effective in achieving these goals, while some improvements could still be made.

First, although using entities to represent topics provides the benefit of not requiring a labeled dataset and being self-explanatory, it is still limited in terms of accuracy. 
Some irrelevant articles still occur in the Article Reviewer stage. 
Also as a result of lacking a human-level performance model that can extract the main characters or events, we did not incorporate the use of temporal information provided by our dataset. 
This is because we can not maintain a consistent thread of storyline evolution from a collection of articles without a human-level performance model.
This is especially critical as our target audience is the general public.

Second, there is still room for improvement to make the system even more user-friendly.
In a multi-stage system like NOVA, provenance~\cite{ragan2015characterizing} can play an important role in helping users keep track of their current status in the narrative structure.
Previous works on provenance mainly focused on analysis for experts, while we believe it can also be useful for general audiences when doing complicated tasks.
Third, the user study has shown that in addition to information exposure, personal beliefs can be influenced by many factors, such as the user's background and social cycle.
The news article datasets can only account for information exposure, leaving the other factors untouched.
Due to the lack of means to collect such data, these factors remain under-explored.
Incorporating these factors into the system design could be a promising direction for future work.

Finally, recent advances in Large Language Models (LLMs) have proven their capabilities to understand user intents, presenting many opportunities to incorporate LLMs into the belief elicitation process. 
For example, LLMs can be used as a chatbot that iteratively facilitates users to externalize beliefs. 
On the other hand, incorporating LLMs for personal belief introduces other challenges and design considerations as well. 
It has been shown that the general public has varying perceptions of the capabilities of LLMs~\cite{zamfirescu2023johnny}. 
How much users will accept the answers given by LLMs and the proper design of human interactions with LLMs in personal belief assessment remains under-researched.
Although NOVA does not incorporate LLMs in the system directly, our design considerations for dividing a complex and provocative task into sub-tasks that are straightforward enough for the general public can inform future research that tries to conduct similar tasks with LLMs. 

\vspace*{-0.1cm}
\section{Conclusion}
We present NOVA, a platform for general audiences to assess their personal beliefs toward media outlets.
NOVA supports personal belief assessment by encouraging belief elicitation from users, contrasting the user's belief with the actual coverage of the media outlets, and finally assisting users in finding evidence to support or refute their beliefs.
Narrative visualization techniques are incorporated to help reduce the mental demand and overload of information to our users.
We design two main visualizations, Sentiment Scatter Plot and Topic Hives to assist users in assessing their personal beliefs.
The visualization and interaction design stresses simplicity and intuitiveness to cater to the general audiences.

To demonstrate and evaluate NOVA we used a subset of our larger dataset that contains articles written during the beginning of the COVID-19 Pandemic in the United States (Feb--Jun 2020). 
From our evaluation, we found NOVA was effective in enabling users to assess their personal beliefs about news outlets.
Even though our participants came from varying backgrounds, age groups, and political affiliations, most of them were able to find evidence to support or refute their beliefs.
For future work, we plan to add provenance support to better assist users in keeping track of the current line of inquiry and offer more context to help lower the overall mental effort. 
Furthermore, our user study shows that information exposure is not the only factor that influences personal beliefs, and a more comprehensive model that accounts for other factors is needed to foster personal belief updating.

\bibliographystyle{abbrv-doi-hyperref}

\bibliography{template}

\vspace*{-1cm}
\begin{IEEEbiography}[{\includegraphics
[width=1in,height=1.25in,clip,
keepaspectratio]{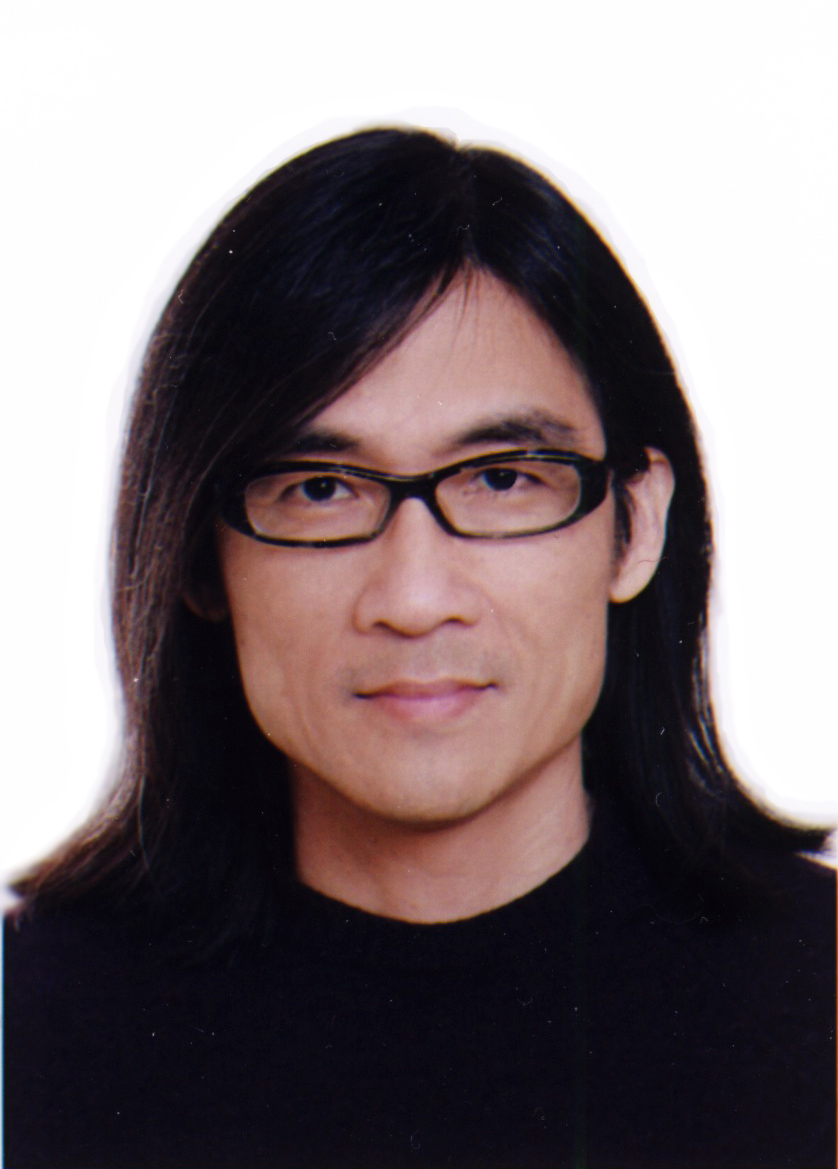}}]
{Kwan-Liu Ma} is a distinguished professor of computer science at the University of California,
Davis, where he leads VIDI Research Group. Professor Ma received his Ph.D. degree in 
computer science from the University of Utah in 1993. His research interests include visualization, computer graphics, human-computer interaction, and machine learning. For his significant 
research accomplishments, Professor Ma has received many recognitions, among others 
the NSF PECASE award in 2000, IEEE Fellow 2012, the IEEE VGTC Visualization Technical
Achievement Award in 2013. the IEEE Visualization Academy in 2019, and ACM Fellow in 2024.
He has served as papers co-chair for SciVis, InfoVis, EuroVis, PacificVis, and Graph Drawing, and on the editorial board of IEEE TVCG (2007-2011) and IEEE CG\&A (2007-2019). 
Professor Ma presently serves on the editorial boards of the Journal of Visual Informatics, the Journal of Computational Visual Media, and ACM Transactions on Interactive Intelligent Systems. Contact him via email: ma@cs.ucdavis.edu.
and research and other interests.
\end{IEEEbiography}

\vspace*{-1cm}
\begin{IEEEbiography}[{\includegraphics
[width=1in,height=1.25in,clip,
keepaspectratio]{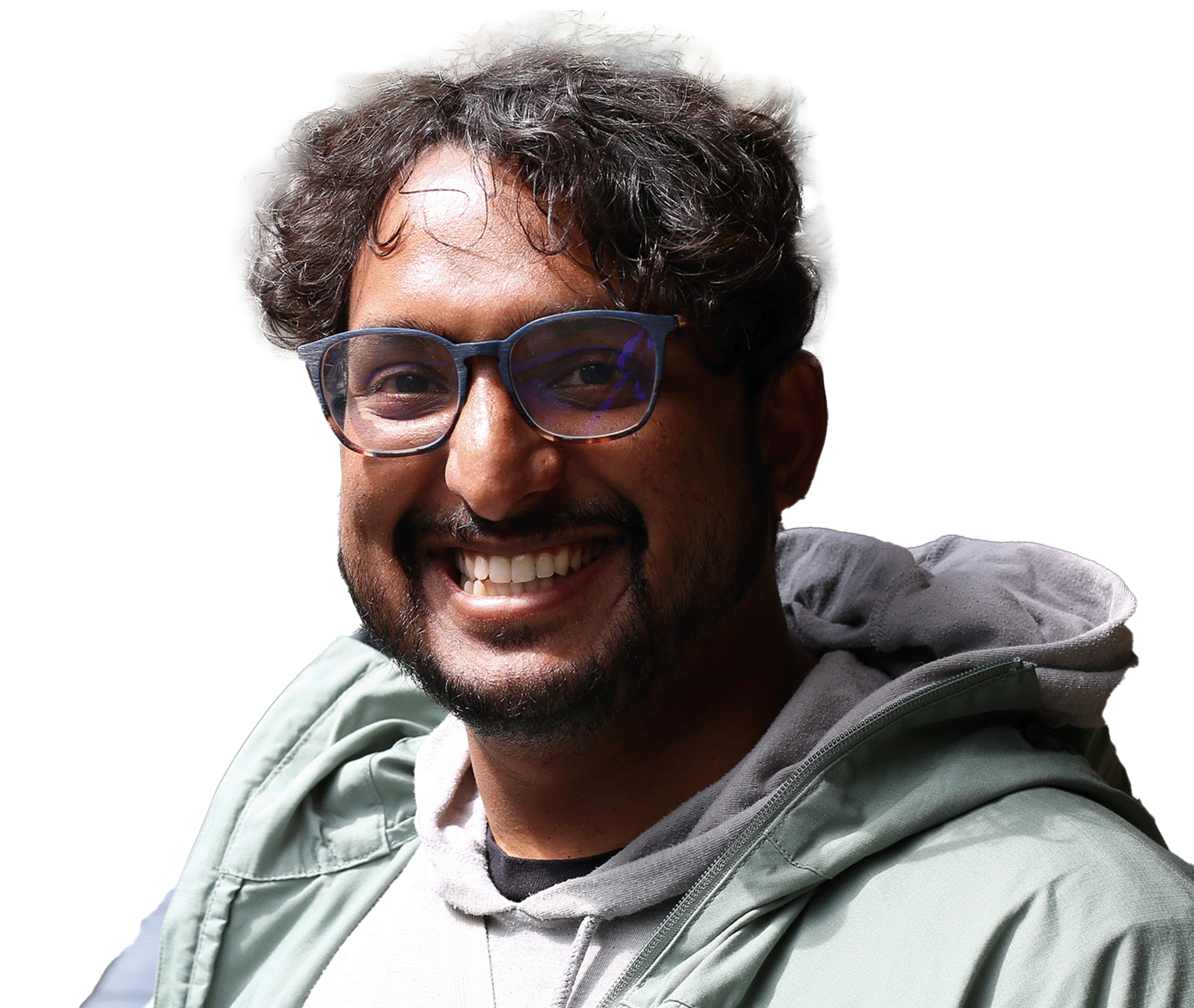}}]{Keshav Dasu} received his Ph.D. degree in computer science from the University of California, Davis in 2023. His research interests include visualization, storytelling, communication, and human-computer interaction.
\end{IEEEbiography}

\vspace*{-1cm}
\begin{IEEEbiography}[{\includegraphics
[width=1in,height=1.25in,clip,
keepaspectratio]{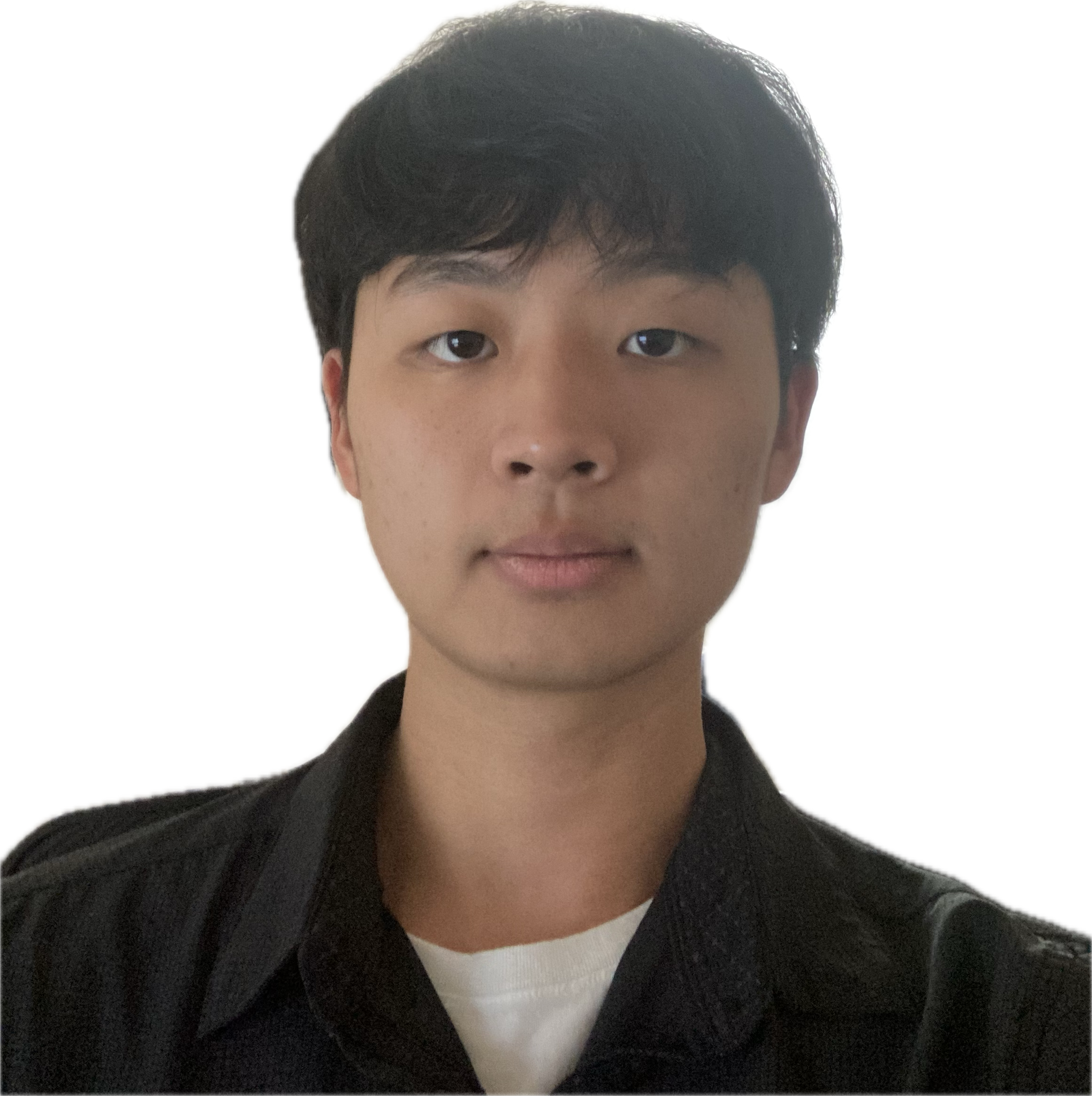}}]{Sam Yu-Te Lee}
received a BS degree in computer science from Nanjing
University in 2020. He is currently a 
PhD student in computer
science at the University of California, Davis. His research interests include text data visualization and analytics.
\end{IEEEbiography}

\vspace*{-1cm}
\begin{IEEEbiography}[{\includegraphics
[width=1in,height=1.25in,clip,
keepaspectratio]{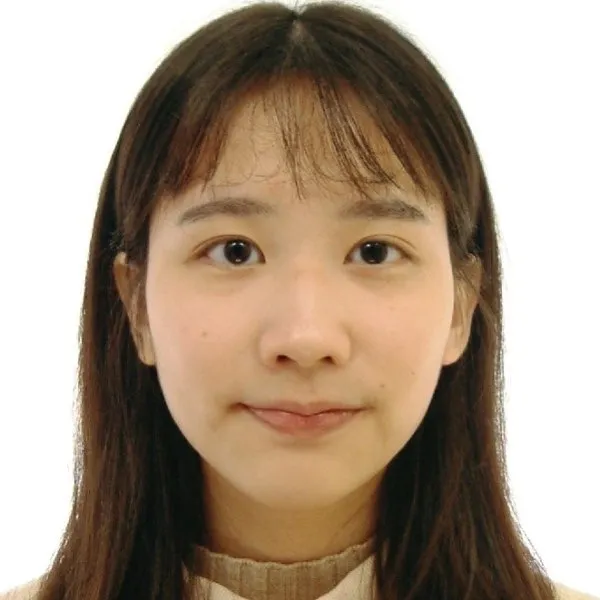}}]{Ying-Cheng Chen}
obtained her bachelor's degree from National Taiwan Normal University in 2022. She currently studies in the Computer Science Master's program at UC Davis, her research focuses on information visualization.
\end{IEEEbiography}


\end{document}